\documentclass[final,5p,twocolum]{elsarticle}

\usepackage{lineno,hyperref}
\usepackage{longtable}
\usepackage{graphics}
\usepackage{graphicx}
\usepackage{epsfig}
\usepackage{amssymb}
\usepackage{amsthm}
\usepackage{lineno}
\usepackage{setspace}
\usepackage{lscape}
\usepackage{xcolor}
\usepackage{lscape}
\usepackage{graphics}
\usepackage{graphicx}
\usepackage{epsfig}
\usepackage{amssymb}
\usepackage{amsthm}
\usepackage{setspace}
\usepackage{lscape}
\usepackage{xcolor}
\usepackage{amsmath}
\usepackage{makecell}

\journal{}

\begin{document}

\begin{frontmatter}
\title{Combined active-passive heat transfer enhancement for a partial superhydrophobic oscillating cylinder}
\author[1]{Ali Rezaei Barandagh}
\author[1]{Adel Rezaei Barandagh}
\author[1]{Jafar Ghazanfarian\corref{cor1}}
\cortext[cor1]{Corresponding author, Tel.: +98(24) 3305 4142.}
\ead{j.ghazanfarian@znu.ac.ir}
\address[1]{Mechanical Engineering Department, Faculty of Engineering, University of Zanjan, P.O. Box 45195-313, Zanjan, Iran.}

\begin{abstract}
Numerical simulation of convective heat transfer over a stationary and transversely oscillating partial super-hydrophobic cylinder has been performed using OpenFOAM libraries. Superhydrophobicity of the cylinder surface has been addressed by means of a partial slip boundary condition. Applying the slip condition to the surface of the stationary cylinder causes the drag and the rms lift coefficients to reduce by 46 and 75 percent, respectively. It also augments the average Nusselt number by 55 percent accompanied by a 21 percent increase of the natural shedding frequency. The partially superhydrophobic cylinder has also been investigated and the effects of slip on different sections of the cylinder surface have been analyzed. Considering the reduction of force coefficients, it is shown that the application of slip over a $135^{\circ}$ segment of the surface is an optimum case, resulting in a 47 and 85 percent decrease of the drag and the rms lift coefficients, respectively. However, the fully superhydrophobic cylinder provides higher heat transfer rates. Regarding the transversely oscillating cylinder, superhydrophobicity extends the primary synchronization region, and also exhibits different wake dynamics behavior compared to the no-slip case. The slip over surfaces also causes the average Nusselt number to become nearly 6 times greater than the no-slip oscillating cylinder at the lock-in condition. Further analysis based on thermal performance index (TPI) proves that a high value of $TPI=6$ can be reached for the superhydrophobic cylinder.
\end{abstract}

\begin{keyword}
\texttt Superhydrophobicity \sep Partial slip \sep Oscillating cylinder \sep Thermal performance index \sep OpenFOAM
\end{keyword}

\end{frontmatter}

\section{Introduction}
Flow and heat transfer over a circular cylinder have been the subject of numerous investigations as a classic example of flow past bluff bodies. Serving as a benchmark problem, it plays an important role in understanding the mechanisms of more complicated applications. Several experimental and numerical studies have been conducted in order to measure and calculate the forces exerted on the body and to help understand complex phenomena such as flow separation and vortex shedding involved in such flows. An unsteady heat transfer from the cylinder is also caused by the unsteady nature of the vortex shedding process. Previous investigations have shown the local heat transfer rate to become maximum near the front stagnation point~\cite{eckert1952u, dennis1968steady}.

Several studies have been carried out both numerically and experimentally regarding an oscillating circular cylinder in a flow stream. The instability mechanism that causes the vortex shedding, namely the Floquete instability, can be controlled by the oscillation of the cylinder in certain ranges of frequency and amplitude by means of a phenomenon known as the lock-in or wake-capture phenomenon. During this range of  synchronization the natural shedding frequency, i.e., the Strouhal number, is lost and the wake oscillates at a frequency equal to the frequency of the body motion. In the past decades, many researchers tried to shed light on this complex fluid-solid interaction. Bishop and Hassan~\cite{bishop1964lift} are among the first researchers who experimentally studied the flow over a transversely oscillating circular cylinder. Their findings showed that the amplitude of the lift and drag forces are comparable with the response of a simple oscillator under the influence of an applied harmonic force. They also found that the average drag force and the lift amplitude increases, while the phase angle between the lift force and the body motion showed a sudden change when the excitation frequency was close to the natural shedding frequency.

Koopman~\cite{koopmann1967vortex} studied the effect of transverse oscillation of the cylinder on the wake geometry, reporting that the cylinder excitation aligned the vortex filaments with the cylinder axis and the lateral spacing of the vortices decreases as the amplitude of oscillation increases. It was further shown that the synchronization phenomenon only occurs above a threshold oscillation amplitude, which attains larger values as the forced oscillation frequency deviates from the natural shedding frequency. The lock-in range was also determined at low Reynolds numbers. Griffin~\cite{griffin1971unsteady} further investigated the influence of a variety of excitation conditions on the cylinder wake geometry from velocity measurements. Through a series of experiments, it was shown that both the amplitude and the frequency of forced oscillations at the lock-in condition affect the formation length that was used as a characteristic parameter. Griffin and Ramberg~\cite{griffin1974vortex} studied the effect of lateral oscillation of a cylinder on vortex shedding. They matched the fluid velocities obtained from experimental measurements with a mathematical model based on the Oseen vortex in order to evaluate the unknown parameters. They have also found an inverse relation between the longitudinal spacing of the vortices and the excitation frequency. In a study by Bearman and Currie~\cite{bearman1979pressure}, the pressure was measured at the lock-in state over a transversely oscillating cylinder at 90 degrees for a broad range of reduced velocities and oscillation amplitudes. They reported a sudden phase jump between the cylinder displacement and the pressure at 90 degrees near the synchronization frequency. Zdravkovich~\cite{zdravkovich1982modification} provided an explanation of the phase jump observed in previous studies and related it to the timing of the vortices being shed with respect to the cylinder displacement. Seo et al.~\cite{jj1} examined a two-phase closed thermosyphon as a passive heat transfer device. They investigated dropwise condensation over a hydrophobic surface. A higher condenser heat transfer coefficient is reported using a polymer-based hydrophobic coating film

Ongoren and Rockwell~\cite{ongoren1988flow} implemented the hydrogen bubble flow visualization technique and showed that for oscillation frequencies lower than the natural shedding frequency, vortices were shed when the cylinder was at its maximum position on the same side. For frequencies of oscillation above the natural shedding frequency, the vortices were shed on the opposite side of the cylinder's maximum position. In an experimental study by Williamson and Roshko~\cite{ch1988vortex}, the effect of oscillation amplitude on the wake formation has been investigated. They classified the vortex shedding patterns based on the number of vortices being shed per oscillation cycle. It was shown that as the amplitude of oscillation increases, the vortex shedding pattern changes from a pair of vortices on one side (2S) to a single vortex on the other side (P+S). They also reported that as the reduced velocity exceeds a certain threshold value for a given oscillation amplitude, the pattern deviates from the 2S mode to a much  more complex form known as the 2P mode, in which two pairs of vortices are shed during each cycle. Furthermore, they related the sudden phase jump to the immediate transition between the 2S and 2P modes  observed in previous studies.

Using the PIV and PTV techniques, Gu et al.~\cite{gu1994timing} carried out an experimental research in which the influence of oscillation frequency on the timing of vortex shedding was approved. In addition, with the oscillation frequency surpassing the natural shedding frequency, two saddle points were observed in the streamline pattern. In a numerical study, Hurlbut et al.~\cite{hurlbut1982numerical} used the finite-difference method to simulate the flow over a cylinder oscillating in both transverse and streamwise directions. Later, the vortex shedding characteristics were numerically investigated by Lecointe and Piquet~\cite{lecointe1989flow} for in-line and lateral oscillations. Meneghini and Bearman~\cite{meneghini1995numerical} solved the oscillatory flow over a circular cylinder using a discrete vortex method at $Re=200$. While determining the lock-in boundary, they extended the oscillation amplitude up to 60 percent of the cylinder diameter with the frequency of oscillation varying around the natural shedding frequency. Kumar et al.~\cite{jj2} conducted pool boiling experiments over heterogeneous wettable surfaces. They printed the polymethyl methacrylate (PMMA) polymer on the plain copper and the hydrophobic polymer on the plain copper and superhydrophilic surface to construct three different cases. They reached enhancement of 98.5\% in heat transfer coefficients.

Anagnostopoulos~\cite{anagnostopoulos2000numerical} provided a numerical solution for the flow over a cylinder that is forced to oscillate transversely using the finite-element technique. The effect of transverse motion of the cylinder on hydrodynamic forces and the wake formation along with the determination of the lock-in boundary were investigated at the Reynolds number of 106. Tang et al.~\cite{tang2017phase} performed a series of numerical simulations for a cylinder subjected to transverse oscillation normal to the incoming flow at $Re=200$. The phase difference between the lift coefficient and transverse displacement, energy transfer between fluid and cylinder as well as the associated vortex shedding modes were addressed and factors that can potentially affect the sign of the phase difference and energy transfer were investigated.

In an experimental research concerning the vibration of a cylinder normal to an airstream by Sreenivasan and Ramachandran~\cite{sreenivasan1961effect}, insignificant variation of the heat transfer coefficient was reported for a maximum velocity amplitude of 0.2. Saxena and Laird~\cite{saxena1978heat} examined an oscillating cylinder in an open water channel and obtained a 60 percent increase in heat transfer coefficient. In experimental studies by Leung et al.~\cite{leung1981heat} and Gau et al.~\cite{gau1999heat}, the heat transfer magnitude was seen to be affected by the oscillation amplitude and frequency. Fu and Tong~\cite{fu2002numerical} investigated the flow and heat transfer characteristics of a transversely oscillating heated cylinder and also found the heat transfer rate to be significantly enhanced in the lock-in regime.

Flow control methods in two categories, namely active and passive techniques, have been designed to control the wake behind bluff bodies such as circular cylinder. Kang et al.~\cite{kang1999laminar} studied the laminar flow over a rotating cylinder in the fully developed stage involving vortex shedding at the Reynolds numbers varying from 47 to 200. The vortex shedding and the wake flow patterns are expected to be modified by the rotation of the cylinder, which may lead to suppressed flow-induced oscillations and amplified lift force. In a numerical study, Ingham and Tang~\cite{ingham1990numerical} examined the flow over a rotating cylinder at $Re<47$ and relatively small non-dimensional rotational speeds ($\alpha<3$). They reported that rotation delays the boundary layer separation, despite the fact that the vortex shedding does not happen in the wake region.
Tang and Ingham~\cite{tang1991steady} considered the steady flow at $Re = 50$ and 100 for the non-dimensional rotational speeds in the interval of $0<a<1$. Nobari and Ghazanfarian determined the flow pattern over a rotating cylinder with transverse oscillations, where the effects of both rotation
and cross flow oscillation on the drag coefficient, the flow field, the lock-in phenomenon, and the wake pattern have been analyzed~\cite[]{nobari2009numerical,ghazanfarian2009numerical}. Flow around a circular cylinder with rotary oscillations has also been investigated in various studies both experimentally, such as Filler et al.~\cite{filler1991response}, and numerically, like Baek and Sung~\cite{baek1998numerical}. Soheibi et al.~\cite{Soheibi} and Amiraslanpour et al.~\cite{Amiraslan} investigated the effect of slotted fins and upstream/downstream splitters, respectively on flow characteristics of an oscillating cylinder.

In comparison to active control methods, passive techniques are generally easier to apply. To name a few, geometric shaping can be used to control the pressure gradient. Fixed mechanical vortex generators and splitters are also considered to be beneficial for separation control~\cite{ozono1999flow}. Many basic ideas in scientific and industrial applications and advancements can be traced back to nature and natural phenomena, one of which is the high water repellency of the Lotus leaf~\cite{barthlott1997purity}. This tendency has inspired the design and construction of superhydrophobic surfaces, which exhibit high water droplet contact angles, generally exceeding 150 degrees~\cite{wang2007definition}. Such behavior results in self-cleaning properties, high corrosion resistance and also drag reduction. Common methods of developing superhydrophobic surfaces either involve surface coating~\cite{nilsson2010novel} or creating certain micro/nano structures and ridges on the surface~\cite[]{belyaev2010effective,oner2000ultrahydrophobic}. The Cassie-Baxter and the Wenzel theories elucidate the relationship between the surface roughness and its wettability. Since the liquid phase completely passes into the roughness grooves, a water-water interface is formed in the Wenzel theory~\cite{wenzel1936resistance}, while in the Cassie-Baxter theory~\cite{cassie1944wettability}, air or some other gases get trapped beneath the liquid, inside the grooves, eventually creating an air-water interface.

A slip boundary has an immense effect on flow pattern, and the drag and lift forces. It is usually characterized by a slip length that is an imaginary distance inside the body starting from the interface, along which the tangential velocity drops to zero. The no-slip boundary condition is assumed to be valid when solving the Navier-Stokes equations in most continuum studies. However, in particular cases such as micro and nano-scale problems and hydrophobic surfaces, this condition may fail~\cite{lauga2007microfluidics}. In their study of flow through thin micro channels, Joesph and Tabeling~\cite{joseph2005direct} directly measured the apparent slip-length on hydrophobic surfaces. In a numerical study, Priezjev et al.~\cite{priezjev2005slip} investigated the effective slip behavior on substrates under shear flow in micro channels with alternating no-slip and shear-free boundary conditions using both continuum and molecular dynamics simulations. You and Moin~\cite{you2007effects} numerically investigated the effect of alternating circumferential bands of the slip and no-slip boundary conditions on the surface of a circular cylinder, which were periodically distributed with different arc lengths. The slip length was 2 percent of the cylinder diameter both in the streamwise and spanwise directions. They reported that the drag force and the root-mean-square of the lift force decrease by as much as 75 percent.

Ou et al.~\cite{ou2007enhanced} used the placement of directional grooves or riblets on the surface to manipulate the flow field and create an effective slip for drag reduction. Lund et al.~\cite[]{hendy2007effective,lund2012calculation} obtained expressions for an effective slip boundary condition in typical cases and extended them to surfaces with periodic roughness. Quere~\cite{quere2005non} and Xue et al.~\cite{xue2012importance} obtained large slip-lengths up to $400$ $\mu$m on super hydrophobic surfaces. The liquid on such surfaces is mostly in contact with air trapped in either structured or unstructured crevices made by the surface treatment. Since the viscosity of air is small, fluid flow over the air-water interface is almost shear-free that reduces the overall drag force. Vakarelski et al.~\cite{vakarelski2012stabilization} showed that this apparent slip effect can be further enhanced by providing a coherent layer of air formed on the surface. The flow around circular cylinders with the slip effect being uniformly distributed on the surface was numerically studied by Legendre et al.~\cite{legendre2009influence}, and the results proved that with increasing the slip length, onset of vortex shedding was delayed and the amount of drag reduction was increased for a given Reynolds number. Park et al.~\cite{jj3} improved the thermal performance of an inclined tube in a two-phase heat exchanger by surface modification techniques. They used the electroplating technique with hydrogen bubbles to create porous microstructures as cavities on a boiling surface and hydrophobic thin films of Teflon. they reported about 107\% enhancement in the boiling heat transfer coefficient.

Slip in unsteady flows has been investigated using the molecular dynamics simulations by Thalakkottor and Mohseni~\cite{thalakkottor2013analysis}. It is found that slip can be determined by both the shear rate and its temporal gradient. Further MD simulations were carried out by Ambrosia et al.~\cite{ambrosia2015static} and Sun et al.~\cite{sun2016hydrophobicity} to obtain equilibrium states of water droplets on groove/ridge textured surfaces using various groove widths and ridge heights. Through a series of experimental investigations on superhydrophobic cylinders with ridges on their surfaces, the Strouhal number and the length of the recirculation region in the wake were shown to be increasing while the rms lift force decreases. It was also reported that superhydrophobicity shifts the onset of vortex shedding towards higher Reynolds numbers~\cite[]{muralidhar2011influence,daniello2013influence}.

In a numerical study by Mastrokalos et al.~\cite{mastrokalos2015optimal}, an increase in the non-dimensional slip-length was shown to have a stabilizing effect on low-Reynolds number flow past a circular cylinder. Kim et al.~\cite{kim2015experimental} determined how the flow separation was affected by rough hydrophobic surfaces. They also investigated the ensuing changes of the vortical structures in the cylinder wake. The effect of superhydrophobicity on viscous and form drag forces was investigated by Huang et al.~\cite{huang2018effect} at different Reynolds numbers (up to 180) and slip-lengths. The viscous drag was found to be dominant at small slip-lengths and Reynolds numbers below 100, while the pressure drag had the main contribution to the total drag at higher Reynolds numbers and slip-lengths. Zeinali et al.~\cite{zeinali2018janus,Behrad2} further investigated the idea of reducing the drag force and the rms lift force using superhydrophobic surfaces. Considering high manufacturing costs and complexities of superhydrophobic surface production, especially at large industrial scales, they introduced the Janus surface concept by means of partially superhydrophobic surfaces. They matched their numerical data to the experimental results obtained by Daniello et al.~\cite{daniello2013influence} for a superhydrophobic cylinder by implementing a partial-slip boundary condition in OpenFOAM codes. The lattice Boltzmann approach can be used to simulate the boiling heat transfer performance on hydrophilic-hydrophobic mixed surfaces~\cite{jj4,jj5}. It is found that an appropriate increase of the contact angle can promote the bubble nucleation on the bottom side and enhances the nucleate boiling on the surface. Also. the interaction between the bubbles nucleated at the corners and the bubbles on the tops of pillars van enhance the departure of the bubbles at corners.

In the present study, the effect of superhydrophobicity will be investigated on flow and heat transfer characteristics of a stationary and transversely oscillating circular cylinder by imposing a partial-slip boundary condition. For the case of a stationary cylinder, the force coefficients, the vortex shedding frequency and the average Nusselt number will be analyzed as well as the local distributions of key parameters such as the pressure coefficient, the skin friction and the Nusselt number. Next, the effects of applying slip on different sections of the cylinder surface will be investigated for the stationary cylinder. Considering the transversely oscillating cylinder, the lock-in boundary will be determined for the fully superhydrophobic cylinder and the effect of slip on the mean drag and the rms lift coefficient will be studied along with the average Nusselt number for different amplitudes and frequencies of oscillation. Also, the wake structure and the vortex shedding modes will be compared to that of a no-slip transversely oscillating cylinder. Finally, using the concept of the thermal performance index (TPI), heat transfer enhancement will be analyzed more thoroughly.

This paper is organized as follows. Section ~\ref{secGovGeo} presents the details of geometry, governing equations and assumptions. The details of numerical technique are discussed in section~\ref{secNum}. Next, the validation of results for various flow and heat transfer characteristics for both the stationary and transversely oscillating cylinders will be presented in section~\ref{secValid}. In section~\ref{secRes} a compendious discussion of the obtained results is presented and lastly, section~\ref{secConc} concludes the paper.

\section{Governing equations and geometry}
\label{secGovGeo}
Considering two-dimensional viscous flow and heat transfer over a transversely oscillating circular cylinder, the governing equations are the continuity, momentum, and energy equations that can be described in the ALE framework as

\begin{equation}
	\frac{\partial{u_{j}}}{\partial{x_{j}}} = 0
\end{equation}
\begin{equation}
	\frac{\partial{u_{i}}}{\partial{t}} + (u_{j} - \psi_{j})\frac{\partial{u_{i}}}{\partial{x_{j}}} = -\frac{1}{\rho}\frac{\partial{p}}{\partial{x_{i}}} + \frac{\partial^2{u_{i}}}{\partial{x_{j}\partial{x_{j}}}}
\end{equation}
\begin{equation}
	\frac{\partial{T}}{\partial{t}} + (u_{j} - \psi_{j})\frac{\partial{T}}{\partial{x_{j}}} = \alpha\frac{\partial^2{T}}{\partial{x_{j}\partial{x_{j}}}}
\end{equation}
where $u_{i}$ is the velocity component, $\psi_{i}$ is the grid velocity component, $\nu$ is the kinematic viscosity, and $\alpha$ is the thermal diffusivity. It can be seen that by setting $\psi_{i}$ to zero, the equations above are reduced to the Eulerian form, and when the fluid velocity component, $u_{i}$, is set equal to the grid velocity component, $\psi_{i}$, the Lagrangian form of the equations are obtained.

A third-type mixed (Robin) boundary condition known as the partial-slip condition has been utilized to represent superhydrophobicty of the cylinder in OpenFOAM codes~\cite{zeinali2018janus}. This boundary condition is as follows
\begin{equation} \label{eqSuperhyd}
u^*_{slip} + (1- \beta) \left( \frac{\partial u^*}{\partial y^*} \right)_{wall} = 0
\end{equation}
where $u^*_{slip}$ is the relative velocity of the fluid at the wall, and $\beta$ is an adjustable coefficient on which a parametric study has been carried out to find a suitable value to close the equations.

The important non-dimensional numbers are the Reynolds number defined based on the free-stream speed and the cylinder diameter
\begin{equation}
	Re = \frac{UD}{\nu}
\end{equation}
the Strouhal number
\begin{equation}
	St = \frac{f_{St}D}{U}
\end{equation}
where $f_{St}$ is the frequency of natural vortex shedding behind the cylinder,
the Prandtl number
\begin{equation}
	Pr = \frac{\nu}{\alpha}
\end{equation}
and the Nusselt number for iso-temperature boundary condition
\begin{equation}
	Nu_{T} = -\frac{D({\partial{T}}/{\partial{n}})}{T_{s} - T_{\infty}}
\end{equation}
where $T_{s}$ and $T_{\infty}$ are the cylinder and free-stream temperatures equal to 330 and 300 K, respectively.
The Nusselt number for the iso-heat flux boundary condition is as follows
\begin{equation}
	Nu_{Q} = -\frac{1}{T^{*}_{s} - T^{*}_{\infty}}
\end{equation}
where $T^{*}$ is the non-dimensional temperature and is equal to
$\frac{T}{D(\partial{T}/\partial{n})}$.
The temperature gradient has been set to 10000 K/m for the case of iso-flux boundary condition. Another important parameter is the thermal performance index (TPI) which is defined as
\begin{equation}
\label{eqTPI}
	\frac{Nu/Nu_{0}}{(C_{d}/C_{d_{0}})^{1/3}}
\end{equation}
where $Nu_{0}$ and $C_{d_{0}}$ represent the initial states of the Nusselt number and the drag coefficient, respectively. This initial state could refer to the stationary case with respect to oscillation or the no-slip state with respect to the superhydrophobic walls. The motion of the cylinder is defined by the following equation
\begin{equation}
y(t) = A\sin(2\pi ft)
\end{equation}
where A and $f$ are the amplitude and frequency of oscillation, respectively. The dimensionless forms of the oscillation amplitude and frequency are $A^* = A/D$ and $F^* = f/f_{St}$. In this paper the values of $A^*$ are equal to 0.2, 0.4, 0.6, 0.8 and $F^*$ is varied in a way that both the non-lock and the lock-in cases occur in the simulations. That is, for the Reynolds number of 200, the non-dimensional oscillation frequency is taken to be 0.5, 0.8, 1, 1.2, 1.5, 2. It should be noted that for the case of the superhydrophobic cylinder, the non-dimensional oscillation frequency of 0.1 has also been taken into account. In addition, the results of the present paper have been generated for the flow of water at $Re = 200$ and $Pr = 7.5$.

\begin{figure}
	\centerline{\includegraphics[width=8.5cm]{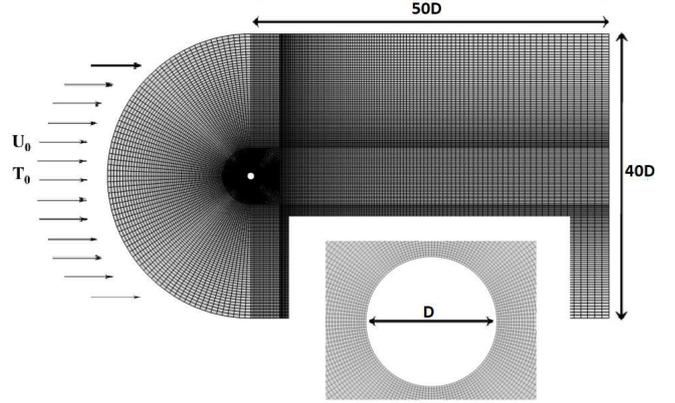}}
	\caption{Details of the C-type structured computational mesh topology and sizes of the domain.}
	\label{fig:Mesh}
\end{figure}

A cylinder with the diameter of D has been placed in a cross-flow. The two-dimensional computational domain consists of an inflow patch which is placed at x = -20D, the top and the bottom boundaries are located at y=$\pm$20D, and the outflow patch is placed at a distance of 50D from the center of the cylinder. Respective boundary conditions are uniform distributions of velocity and temperature, zero-gradient condition for pressure at the inlet; zero fixed-value for pressure, zero-gradient for velocity and temperature at the outlet; and both the no-slip and the partial-slip along with the iso-temperature and iso-flux boundary conditions over the cylinder surface. A two-dimensional structured mesh has been generated with higher mesh density near the cylinder in order to accurately resolve the boundary layer behavior close to the cylinder surface. Details of the computational domain and the mesh as well as the aforementioned boundary conditions can be seen in Fig.~\ref{fig:Mesh}.

\section{Numerical method}
\label{secNum}
The finite-volume method has been used through the open source computational fluid dynamics code, i.e., OpenFOAM. OpenFOAM is a set of C++ libraries and tools aimed at solving the problems of continuum mechanics, specially CFD applications, by means of several discretization techniques and numerical solvers, as well as various pre/post-processing utilities. Laminar incompressible flow over a circular cylinder with transverse oscillation has been solved using OpenFOAM codes. To do so, a dynamic-mesh flow and energy solver, called ThermalPimpleDyMFoam,  previously developed by Ghazanfarian and Taghilou~\cite{ghazanfarian2017active} has been used.

Regarding the discretization of the time derivatives and the gradient terms, the Euler method and the Gauss-linear scheme were used, respectively. For the diffusion terms, a second-order Gauss-linear method was implemented and a second-order upwind scheme was used to discretize the convective terms. For the purpose of coupling the pressure and velocity, the PIMPLE algorithm was used. This algorithm can be thought of as a combination of the PISO and SIMPLE‌ algorithms, all being iterative methods. It should be mentioned that a better stability is gained in PIMPLE over PISO, especially when dealing with large time-steps. The under relaxation factors were set to 0.5, 0.7, 1 for pressure, momentum and energy equations, respectively. Furthermore, the convergence tolerance for pressure was $10^{-7}$ and $10^{-9}$ for other parameters. The Courant number was also kept less than unity throughout the simulations.

\section{Validation and verification}
\label{secValid}
In this section, the obtained results are compared
to the reported data in previous studies to
ensure the accuracy of the present simulations. First, the mesh and time-step
size independence tests are carried out, and then appropriate mesh/time-step sizes for the simulations are suggested.
Next, the acquired results for flow and heat transfer
around the fixed and the oscillating cylinder will be compared with the available data.

\subsection{Mesh/time-step size independence tests}
\begin{figure*}
	\centerline{\includegraphics[width=\textwidth]{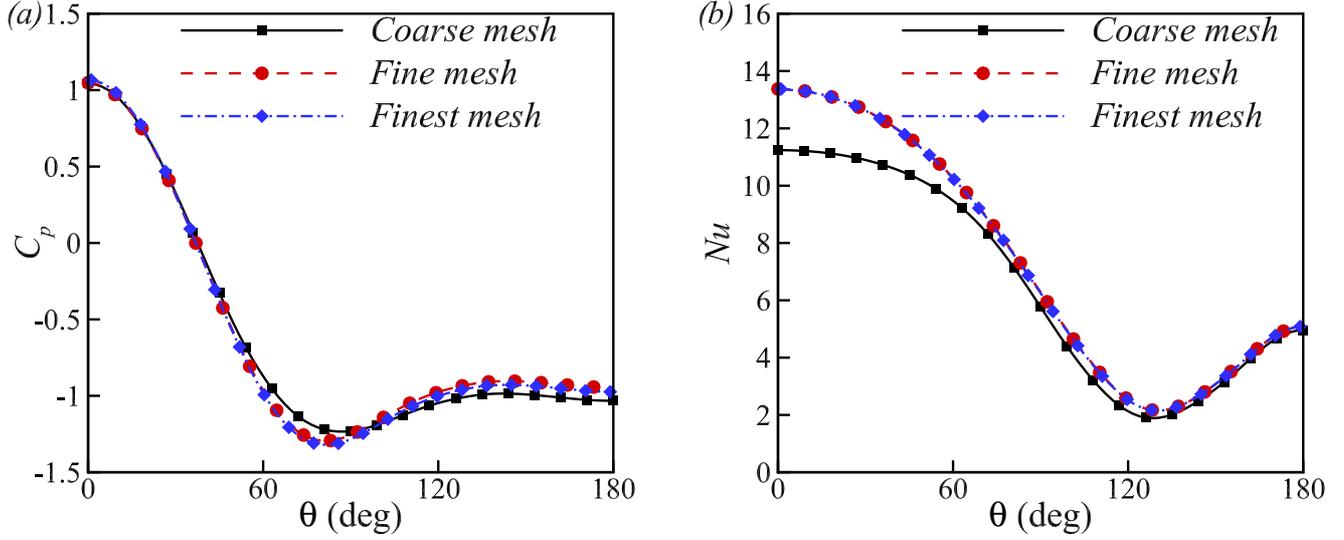}}
	\caption{Local variation of the time-averaged $(a)$ pressure coefficient, and $(b)$ Nusselt number for three grid resolutions at $Re = 200$ and $Pr = 0.71$.}
	\label{fig:LocalCpNuMeshDep}
\end{figure*}

In order to achieve grid-independent results, a set of simulations have been performed on three different computational meshes at $Re = 200$. Table~\ref{tab:MeshInd} summarizes the details of mesh topologies and corresponding results. As can be seen, the difference between the obtained values for the drag coefficient, the Strouhal number, and the average Nusselt number is less than 1 percent for the fine and finest cases. Also, Fig.~\ref{fig:LocalCpNuMeshDep} shows the variation of pressure coefficient and Nusselt number on the surface of the cylinder for three grid resolutions. It can be observed that the difference between the fine and finest cases is less than 3 percent for both the pressure coefficient and Nusselt number distributions. Therefore, the fine grid is sufficient to perform simulations.
\begin{table*}
	\begin{center}
		\caption{Results of the mesh independence test for the flow around the fixed cylinder at $Re = 200$ and $Pr = 0.71$.}
		\def~{\hphantom{0}}
		\begin{tabular}{lcccccc}
			\hline\hline
			& Grid & $C_{d}$   &   $St$ & $Nu$ & No. of cells & No. of cells on the cylinder  \\[3pt]\hline
			\hline
			& Coarse & $1.419$ & $0.200$ & $7.739$ & 70178 & 158\\
			\hline
			& Fine & $1.355$ & $0.193$ & $7.444$ & 118736 & 160\\
			\hline
			& Finest & $1.358$ & $0.193$ & $7.508$ & 155134 & 170\\
			\hline\hline
		\end{tabular}
		\label{tab:MeshInd}
	\end{center}
\end{table*}

The numerical predictions were also investigated with respect to three different time-step sizes. The values of the mean drag coefficient, the Strouhal number, and the average Nusselt number have been computed as listed in Tab.~\ref{tab:TimeInd}. The results for the normalized time-step size of 0.005 is very close to the results generated with $\Delta t^*=0.0025$. Therefore, the time-step size of 0.005 is found to be sufficient to generate results.
\begin{table}
	\begin{center}
		\caption{Results of the time-step size independence test for the flow around the fixed cylinder at $Re = 200$ and $Pr = 0.71$.}
		\def~{\hphantom{0}}
		\begin{tabular}{lcccc}
			\hline\hline
		    & $C_{d}$   &   $St$ & $Nu$ & Normalized time step ($\Delta t^{*}$) \\[3pt]\hline
			\hline
			& $1.350$ & $0.191$ & $7.654$ & 0.01\\
			\hline
			& $1.355$ & $0.193$ & $7.444$ & 0.005\\
			\hline
			& $1.356$ & $0.193$ & $7.447$ & 0.0025\\
			\hline\hline
		\end{tabular}
		\label{tab:TimeInd}
	\end{center}
\end{table}

\begin{figure*}
\centerline{\includegraphics[width=\textwidth]{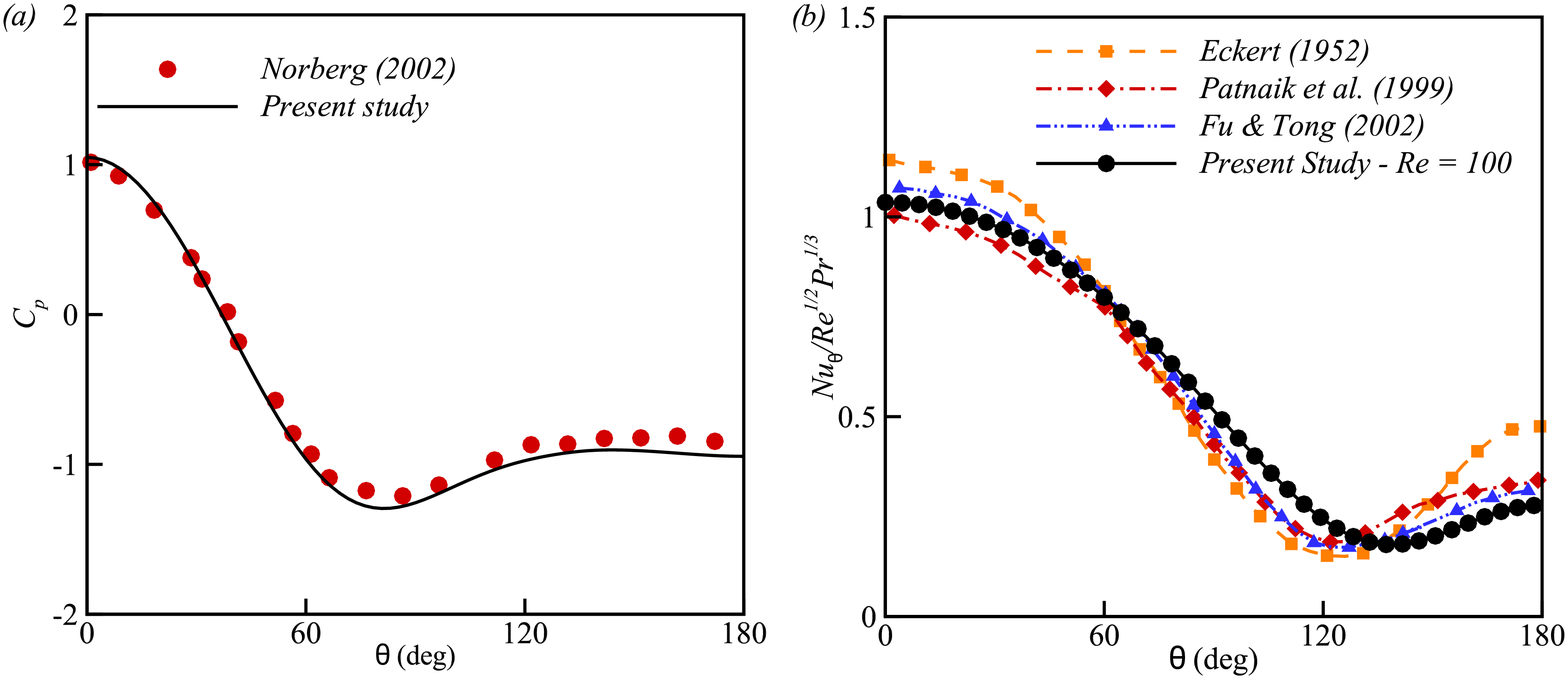}}
\caption{Local variation of the time-averaged $(a)$ pressure coefficient, data obtained from Norberg~\cite{norberg2002pressure} at $Re = 200$, and $(b)$ Nusselt number, data obtained from Eckert \cite{eckert1952u}, Fu and Tong~\cite{fu2002numerical} and Patnaik et al.~\cite{patnaik1999numerical} at $Re = 100$.}
\label{fig:LocalCpNuValidStat}
\end{figure*}

In order to validate the data obtained in the present study for the case of fixed cylinder, the drag and lift coefficients ($C_{d}$ and $C_{l}$, respectively), the Strouhal number and the average value of the Nusselt number have been compared with the previous experimental and numerical results available in the literature at $Re=200$ and $Pr = 0.71$ in Tab.~\ref{tab:ValidationStationary}. As can be seen, the present computations are in good agreement with those calculated before, showing an overall difference of 5 percent. The local variation of the pressure coefficient and the Nusselt number have been illustrated in Fig.~\ref{fig:LocalCpNuValidStat}(a) and (b), respectively. The difference between the current calculations and the previous data is under 15 percent. Note that the local Nusselt number around the cylinder has been computed at $Re=100$ in order to be comparable with the results obtained in the previous studies.

\begin{table*}
\begin{center}
\caption{The mean drag coefficient, amplitude of the lift coefficient, the Strouhal number, and the average Nusselt number for flow over a stationary circular cylinder at $Re = 200$ and $Pr = 0.71$.}
\def~{\hphantom{0}}
\begin{tabular}{lcccc}
\hline\hline
Literature data & $C_{d}$   &   $C_{l}$ & $St$ & $Nu$\\[3pt]
\hline\hline
Persillon \& Braza \cite{persillon1998physical}   & 1.345 & ~~0.7~ & 0.204 & -\\
\hline
Liu \emph{et al.} \cite{liu1998preconditioned}   & 1.31$\pm$0.05 & ~~0.69 & 0.192 & -\\
\hline
Qu \emph{et al.} \cite{qu2013quantitative} & 1.32$\pm$0.01 & ~~0.66 & 0.196 & -\\
\hline
Kim \& Choi \cite{kim2019lock}   & 1.35$\pm$0.05 & ~0.7 & 0.197 & -\\
\hline
Churchill \& Bernstein \cite{churchill1977correlating} & - & - & - & 7.227\\
\hline
Bergman \emph{et al.} \cite{bergman2011fundamentals} & - & - & - & 7.453\\
\hline
Present study & 1.355 & 0.693 & 0.193 & 7.444\\
\hline\hline
\end{tabular}
\label{tab:ValidationStationary}
\end{center}
\end{table*}

\begin{figure*}
\centerline{\includegraphics[width=\textwidth]{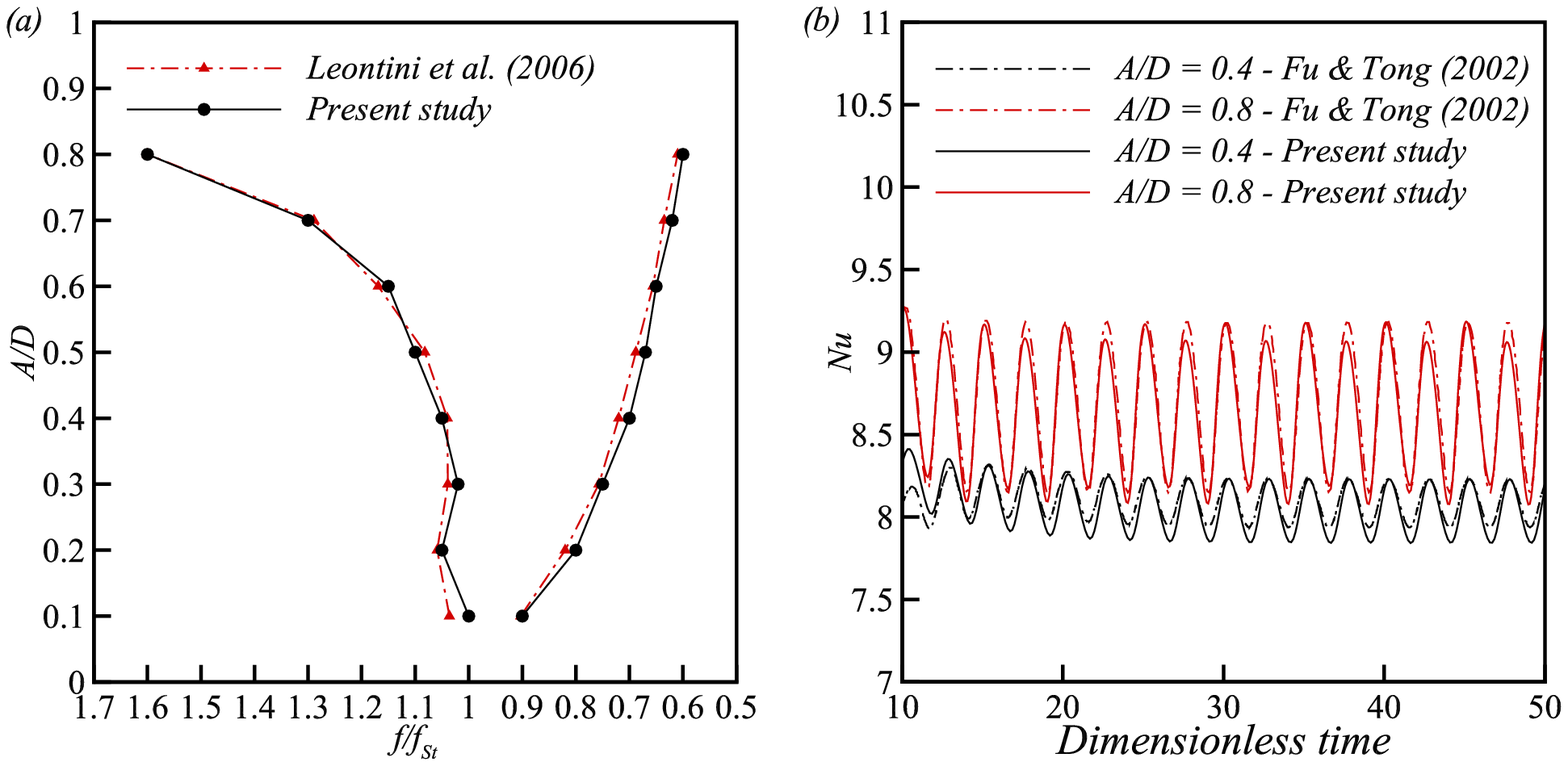}}
\caption{Comparison of $(a)$ primary synchronization region with the map of Leontini et al.~\cite{leontini2006wake}, and $(b)$ temporal variation of the Nusselt number at $f/f_{St}=1$ for two different oscillation amplitudes of $A/D=0.4$ (the lower curves) and $A/D=0.8$ (the upper curves), data obtained from Fu and Tong~\cite{fu2002numerical}.}
\label{fig:ValidLockInNu}
\end{figure*}

\begin{figure*}
\centerline{\includegraphics[width=\textwidth]{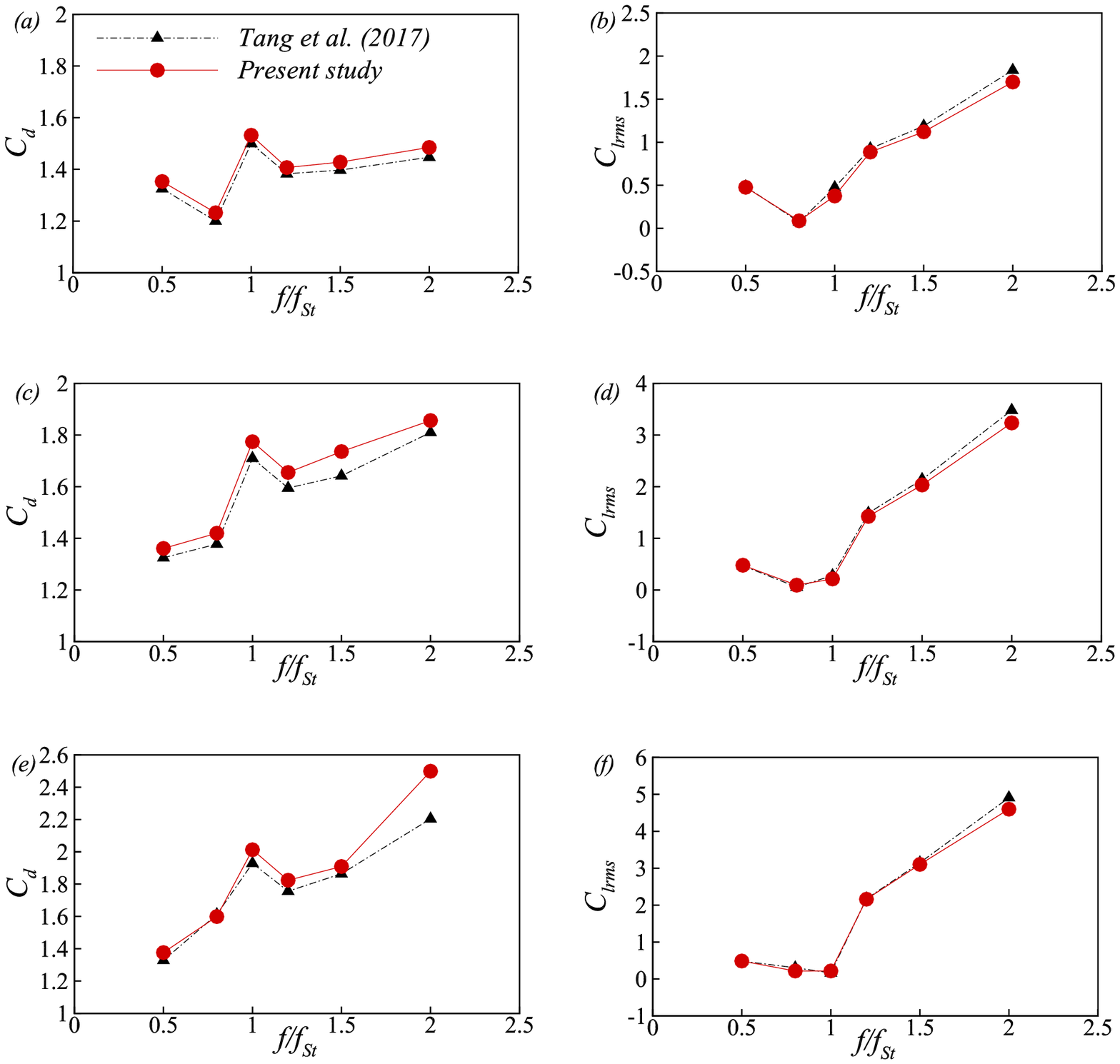}}
\caption{Comparison of the mean drag coefficient and the rms lift coefficient with the results of Tang et al.~\cite{tang2017phase}, $A^*$ is equal to 0.2 for the first row, 0.4 for the second row, and 0.6 for the third row.}
\label{fig:TangValid}
\end{figure*}

Next, the lock-in boundary, the time-history of the Nusselt number, the mean value of the drag and the rms lift coefficients have been verified by the previous data at hand for the case of transversely oscillating cylinder. The primary synchronization range, i.e., $f/f_{St} \simeq 1$, has been determined and compared with the map of Leontini et al.~\cite{leontini2006wake} in Fig.~\ref{fig:ValidLockInNu}$(a)$, showing an overall deviation of 3 percent from the previously obtained results. Figure~\ref{fig:ValidLockInNu}$(b)$ shows the time-history of the Nusselt number for the oscillating cylinder case at two different oscillation conditions. The present values have been verified by the results of Fu and Tong~\cite{fu2002numerical}, indicating a good agreement between the two computations with a maximum error of 3 percent.
The mean value of the drag coefficient and the rms lift coefficient have also been compared with the results obtained by Tang et al.~\cite{tang2017phase} in Fig.~\ref{fig:TangValid} for various frequency ratios and oscillation amplitudes where a total difference of 10 percent is seen between the computed results and the previous data.

\section{Results}
\label{secRes}
In this section the results of applying superhydrophobicity on the flow and heat transfer characteristics of the stationary and transversely oscillating cylinder will be provided. For the case of a stationary cylinder, mean force coefficients and heat transfer rates will be analyzed along with the local distributions of flow and heat transfer parameters. Also, the application of slip along different segments of the cylinder surface will be investigated for the stationary case. Next, the lock-in boundary and average drag and lift coefficients will be studied for the oscillating cylinder. Furthermore, the effect of superhydrophobicity on the distribution of vorticity and vortex shedding modes will be investigated as well as heat transfer rates and temperature contours. Finally, the relative variations of drag coefficient and Nusselt number will be analyzed by means of the $Nu/C_{d}$ ratio and thermal performance indices.

\subsection{Stationary cylinder}
\begin{figure*}
\centerline{\includegraphics[width=14cm,height=14cm,keepaspectratio]{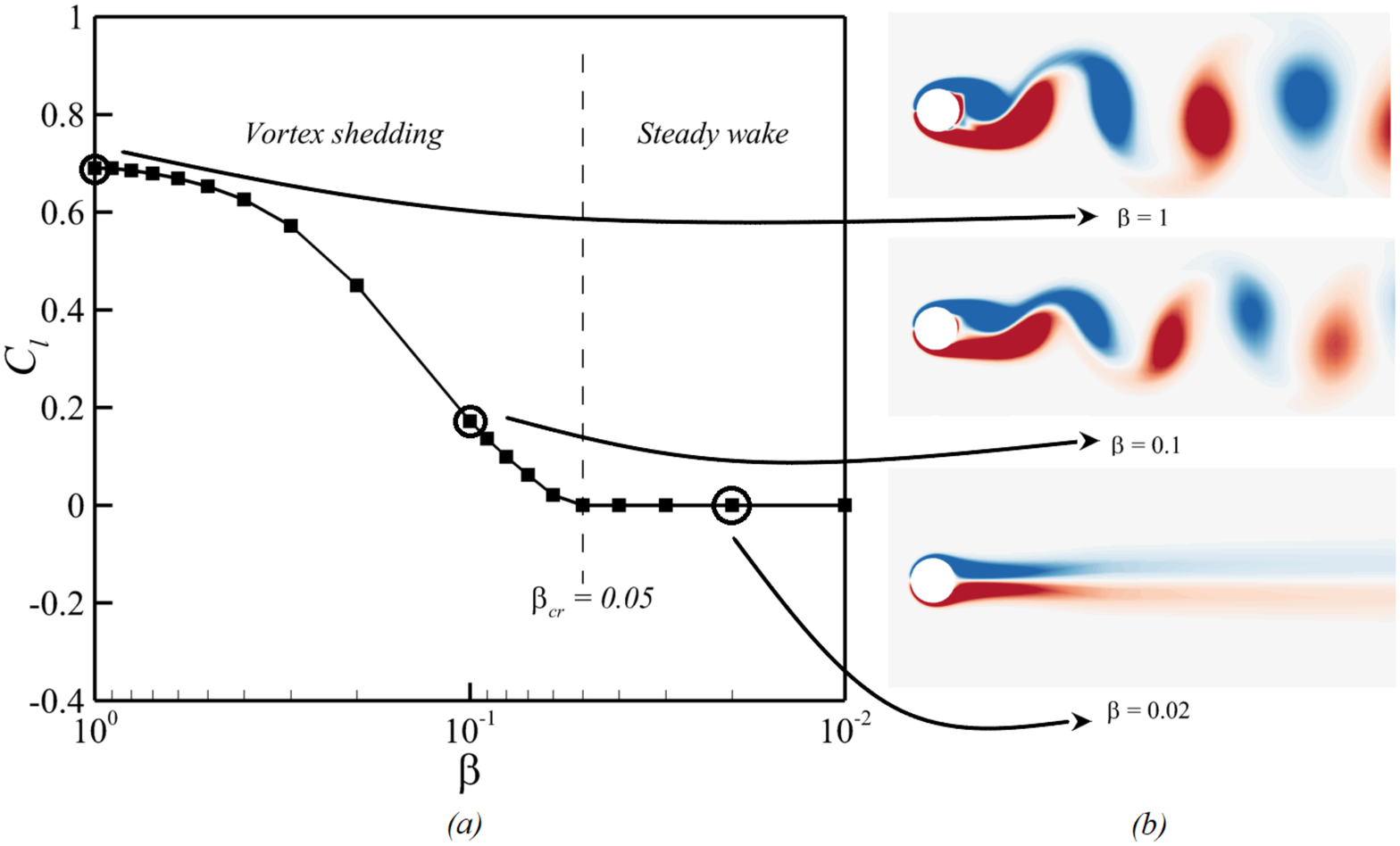}}
\caption{(a) Variation of the lift coefficient amplitude against the slip coefficient in the steady and vortical regimes, and (b) comparison of the wake structure for three cases of $\beta =$ 0.02, 0.1, 1.}
\label{fig:Map}
\end{figure*}

In order to investigate the influence of the partial-slip condition on the flow and heat transfer characteristics of the stationary cylinder, we first examine the effect of varying the slip coefficient. Note that $\beta=1$ corresponds to the no-slip condition, hence a decrease in $\beta$ in the range $0<\beta<1$ leads to an increased amount of slip. Similar to the results reported by Legendre et al.~\cite{legendre2009influence}, there exists a threshold value for the slip coefficient below which vortex shedding does not occur and the wake remains steady. This trend can be seen in our data illustrated in Fig.~\ref{fig:Map}$(a)$ as the lift amplitude decreases with increasing the amount of slip and drops off to zero at $\beta_{cr}=0.05$. In order to shed light on such behavior, Fig.~\ref{fig:Map}$(b)$ demonstrates the vorticity contours for three cases of $\beta = 0.02, 0.1, 1$. It can be seen that the wake turns to steady state for the case of $\beta=0.02$, and the superhydrophobic condition causes the vortices to become more stretched alongside the streamwise direction compared to the no-slip cylinder ($\beta=1$).

\begin{figure*}
\centerline{\includegraphics[width=\textwidth]{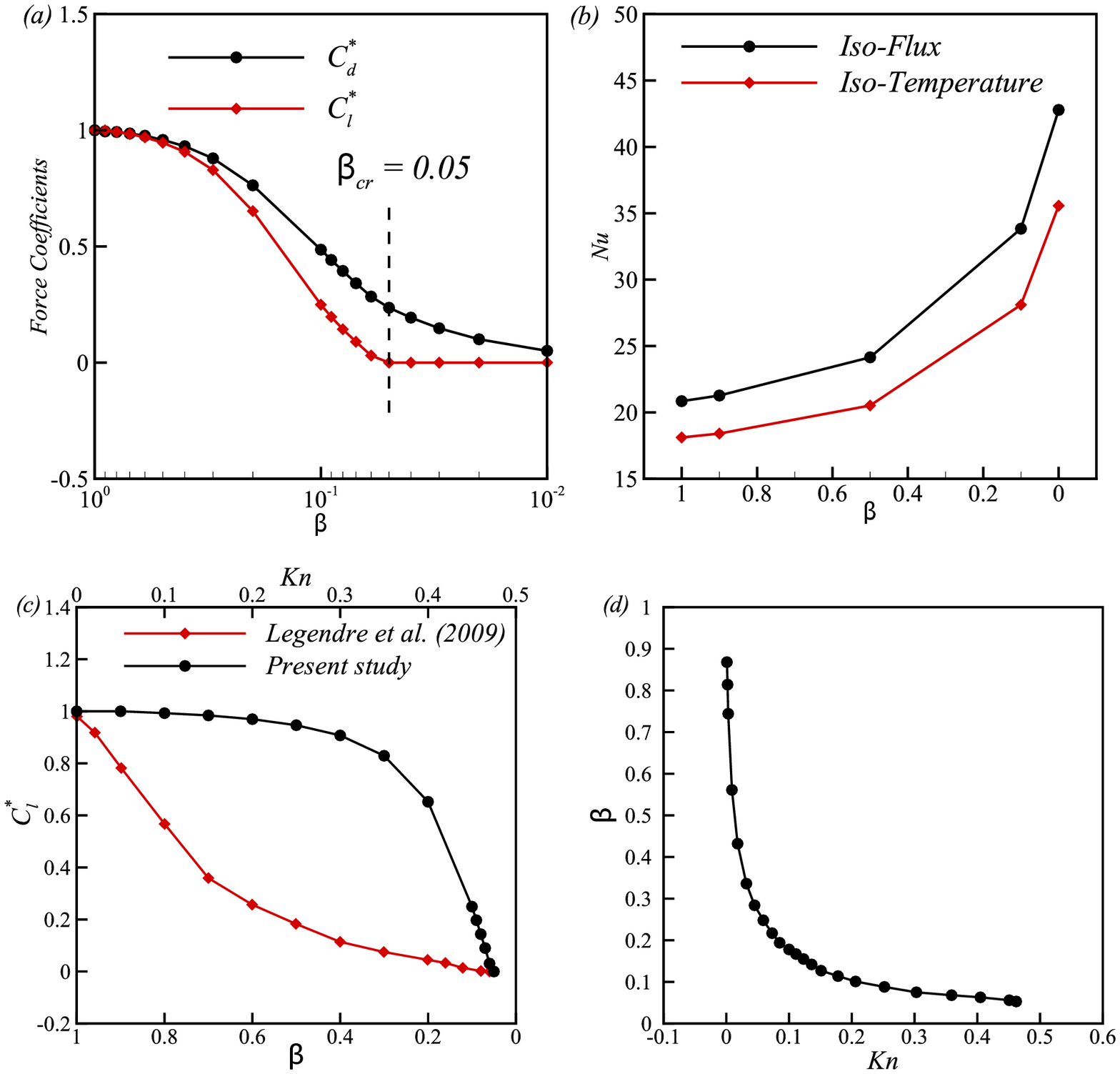}}
\caption{Variation of (a) the force coefficients, (b) the Nusselt number against the slip coefficient, (c) the normalized lift amplitude, and (d) the corresponding values of slip coefficient against the non-dimensional slip length for the case of stationary cylinder, data obtained from Legendre et al.~\cite{legendre2009influence}.}
\label{fig:ParamStudyBoth}
\end{figure*}

Figure~\ref{fig:ParamStudyBoth}$(a)$ presents variations of the force coefficients as $\beta$ attains different values. It is found that the values of the normalized mean drag and lift amplitude decrease as $\beta$ attains lower values, showing that the force coefficients can be reduced up to 90 percent when the slip coefficient reaches zero. It should be noted that the values of drag coefficient have been normalized with respect to both the reference no-slip and shear-free values, such that $C_{d}^{*} = (C_{d} - C_{d}(0))/(C_{d}(1) - C_{d}(0))$, where $C_{d}(0)$ and $C_{d}(1)$ are the values of drag coefficient for the shear-free and no-slip conditions, equal to $0.131$ and $1.355$, respectively. Also, $C_{l}^{*} = C_{l}/C_{l}(1)$, where $C_{l}(1)$ is the no-slip lift amplitude, equal to $0.690$. Variation of the average Nusselt number is also shown in Fig.~\ref{fig:ParamStudyBoth}$(b)$ for iso-temperature and iso-flux boundary conditions. It is obvious that both curves exhibit the same behavior under different slip conditions. As a result, it can be seen that the heat transfer rate is enhanced for about 100 percent with applying increased slip on the surface. It should be noted that only the iso-temperature boundary condition has been considered hereinafter.

In order to further examine the behavior of different flow parameters for the superhydrophobic cylinder, the link between the slip coefficient and the slip length needs to be pointed out. Figure.~\ref{fig:ParamStudyBoth}$(c)$ shows the variation of the normalized lift amplitude alongside with the results of Legendre et al.~\cite{legendre2009influence}, where $Kn$ is the non-dimensional slip length. The relationship between the slip length and the slip coefficient could be attained using this figure, such that the corresponding values of the mentioned parameters are plotted against each other in Fig.~\ref{fig:ParamStudyBoth}$(d)$. This diagram could be used to associate the value of slip length to a desired slip coefficient. For the purpose of determining the accuracy of this diagram, $Kn = 0.2$ is selected, where the corresponding value of the slip coefficient is found to be $\beta = 0.1$. The values of the normalized drag coefficient and lift amplitude as well as the normalized Strouhal number ($St^{*} = St/St(1)$, where $St(1)$ is the Strouhal number for the no-slip case, i.e. $0.193$) are stated in Tab.~\ref{tab:KnBRes}. It can be seen that the overall difference between the results is less than 5 percent.

\begin{table}
\begin{center}
\caption{Results of the corresponding slip length and slip coefficient values for the flow around the fixed superhydrophobic cylinder.}
\def~{\hphantom{0}}
\hspace{-1.0 cm}
\begin{tabular}{lcccc}
\hline\hline
&   &  $C_{d}^{*}$   & $C_{l}^{*}$ & $St^{*}$ \\[3pt]\hline
\hline
& Legendre et al.~\cite{legendre2009influence}, $Kn = 0.2$ & $0.436$ & $0.257$ & $1.228$\\
\hline
& Present study, $\beta = 0.1$ & $0.486$ & $0.249$ & $1.207$\\
\hline\hline
\end{tabular}
\label{tab:KnBRes}
\end{center}
\end{table}

We choose $\beta = 0.1$ to further examine the effect of superhydrophobicity on various parameters of the flow. Different flow and heat transfer parameters are shown in Tab.~\ref{tab:DragComp} for the no-slip and superhydrophobic cylinders. Results in the table indicate that the total drag coefficient has decreased by 46.2 percent. Slip also causes an almost 50 percent decrease in the form drag accompanied by a 40 percent reduction in the friction drag. The amplitude and rms of the lift coefficient have been pronouncedly suppressed both by an amount of 75 percent. The Nusselt number is also 55 percent higher in the case of the superhydrophobic cylinder. It can also be seen that the ratio of $Nu/C_{d}$ goes up significantly by applying the slip condition, showing a 189.59 percent increase. Lastly, the separation angle is 20.51 percent higher for the superhydrophobic case.
\begin{table*}
\caption{Values of the form drag coefficient, the viscous drag coefficient, the total drag coefficient, the amplitude and rms of the lift coefficient, and the Nusselt number obtained for the stationary cylinder regrading the no-slip and the superhydrophobic conditions ($\beta = 0.1$).}
\hspace{-0.0 cm}
\begin{tabular}{ccccccccc}
\hline\hline
& $C_{dp}$   &   $C_{df}$ & $C_{d}$ & $C_{l}$ & $C_{lrms}$ & $Nu$ & $Nu/C_{d}$ & $\theta _{sep}$ \\[3pt]
\hline\hline
The no-slip case & 1.077 & ~~0.278~ & 1.355 & 0.690 & 0.487 & 18.114 & 13.368 & $111.640^{\circ}$ \\
\hline
The superhydrophobic case  & 0.548 & ~0.178 & 0.726 & 0.172 & 0.121 & 28.106 & 38.713 & $134.545^{\circ}$ \\
\hline
Percentage of increase/decrease   & 49.11 & ~35.97 & 46.42 & 75.07 & 75.15 & 55.16 & 189.59 & 20.51 \\
\hline\hline
\end{tabular}
\label{tab:DragComp}
\end{table*}

\begin{figure*}
\centerline{\includegraphics[width=0.8\textwidth]{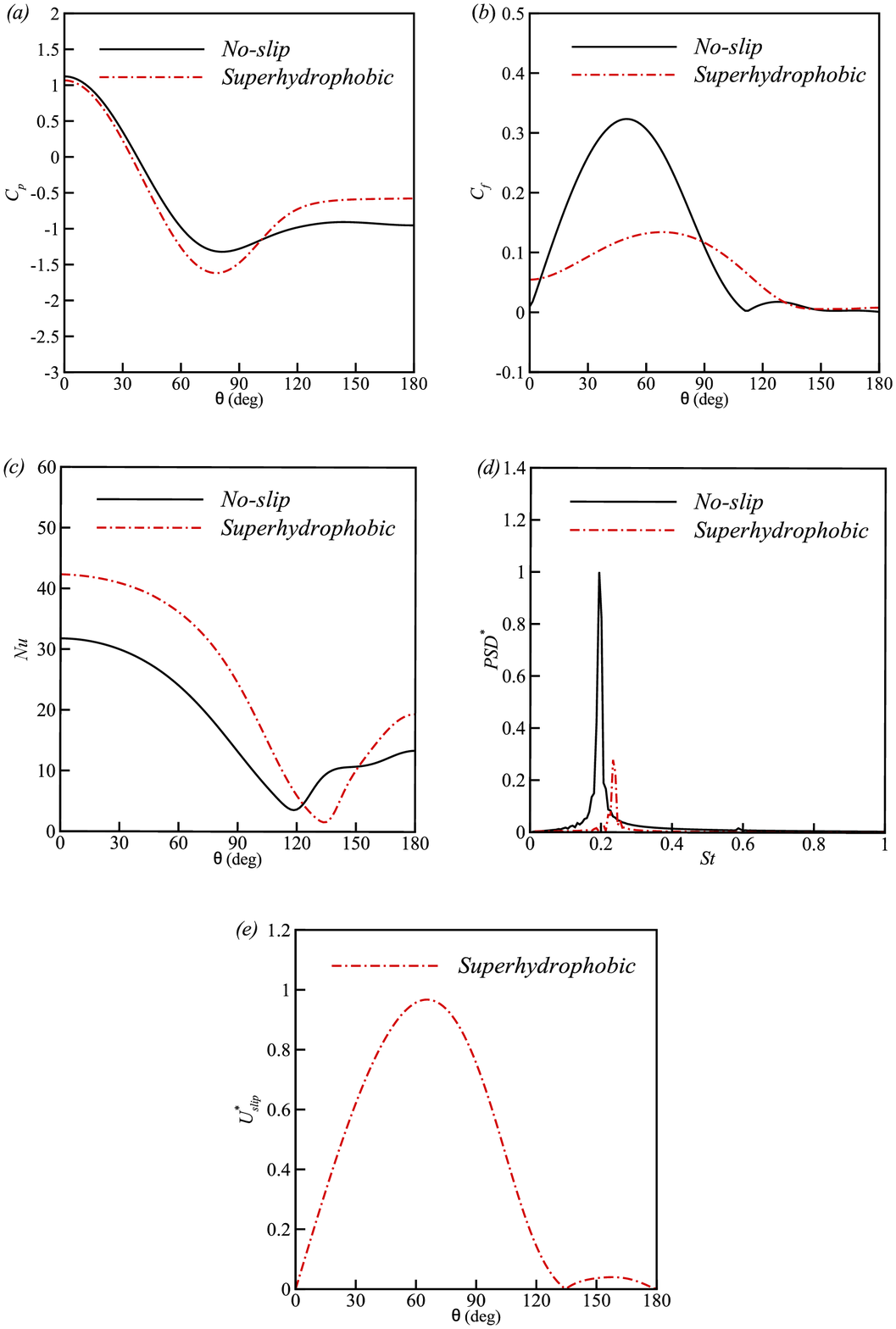}}
\caption{Local variation of the time-averaged (a) pressure coefficient, (b) skin friction coefficient, (c) Nusselt number, (d) fast Fourier transformation of the lift coefficient, and (e) the slip-velocity for the no-slip and superhydrophobic stationary cylinders.}
\label{fig:CpCfResultNuFFTResult}
\end{figure*}

\begin{figure*}
\centerline{\includegraphics[width=\textwidth]{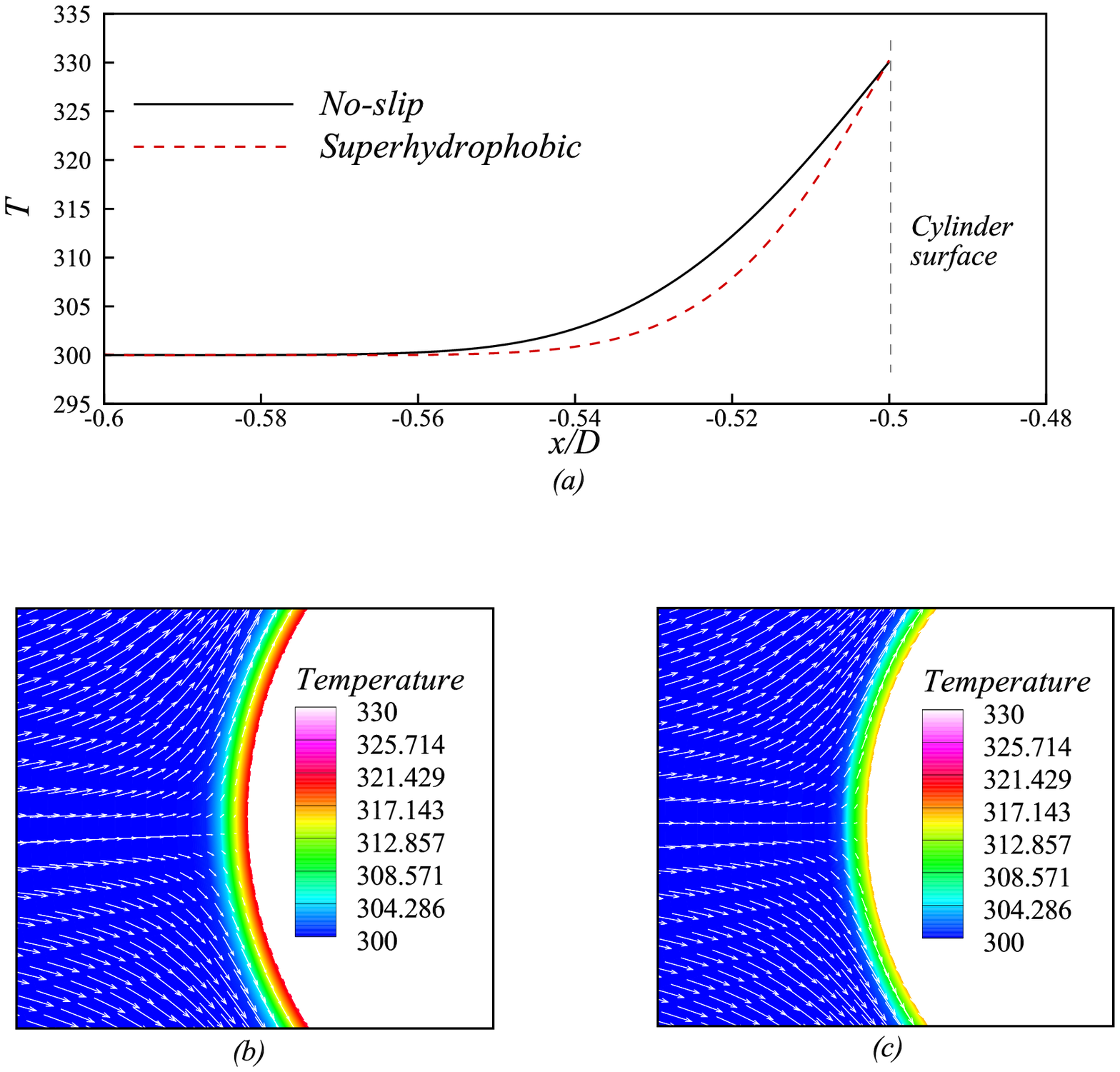}}
\caption{(a) The temperature profile (kelvin) along the streamwise direction and close-up views of (b) the no-slip and (c) the superhydrophobic cylinder at the front stagnation point.}
\label{fig:TCloseUp}
\end{figure*}

The local distribution of form drag and the skin friction coefficient are also shown in Fig.~\ref{fig:CpCfResultNuFFTResult}$(a)$ and $(b)$, respectively. As can be seen, the pressure difference between the front and rear stagnation points shows a 24 percent decrease for the superhydrophobic cylinder, which is responsible for the pressure drag reduction mentioned above. Furthermore, superhydrophobicity significantly reduces the skin friction coefficient over most of the cylinder surface, showing a 65 percent decrease of its maximum value. Figure~\ref{fig:CpCfResultNuFFTResult}$(c)$ illustrates the local variation of the Nusselt number for both the no-slip and superhydrophobic cylinders. It is clear that heat transfer is enhanced near the front and rear stagnation points. For both cases, the highest value of the Nusselt number is attained at $\theta=0^{\circ}$, which is the front stagnation point, and the lowest value lies between the  separation point and the rear stagnation point. Therefore, it can be observed that applying the slip condition increases the heat transfer rate throughout most of the cylinder surface, starting from the front stagnation point up until where separation occurs. This matter is further analyzed for the front stagnation point in Fig.~\ref{fig:TCloseUp}$(a)$, where the temperature profile has been plotted along the streamwise direction, starting from the surface of the cylinder into the upstream field. As is depicted, the temperature gradient is higher for the superhydrophobic cylinder at the front stagnation point which explains the larger value of $Nu$ at this position. This trend stems from higher rates of convection due to slip in this region, as shown by the velocity vectors in Fig.~\ref{fig:TCloseUp}$(b)$ and $(c)$. Spectral analysis of the lift coefficient is also carried out by means of the fast Fourier transform and the result is reported in Fig.~\ref{fig:CpCfResultNuFFTResult}$(d)$. The figure proves that superhydrophobicity increases the dimensionless vortex shedding frequency, i.e., the Strouhal number by almost 21 percent, from 0.193 to 0.233. This increasing trend is in accordance with the results reported in previous studies~\cite{zeinali2018janus}. It is seen that the maximum value of the normalized power density reduces remarkably for the superhydrophobic cylinder, which shows a 70 percent decrease. Finally, Fig.~\ref{fig:CpCfResultNuFFTResult}$(e)$ depicts the variation of the mean dimensionless slip-velocity along the surface of the cylinder, which has been normalized using the free-stream velocity. It is shown that the amount of $U^{*}_{slip}$ rises to its maximum value at $\theta$ around $65^{\circ}$, and after falling down to zero at the separation point, remains extremely low in the wake region.

\begin{figure}
\centerline{\includegraphics[width=8.5cm]{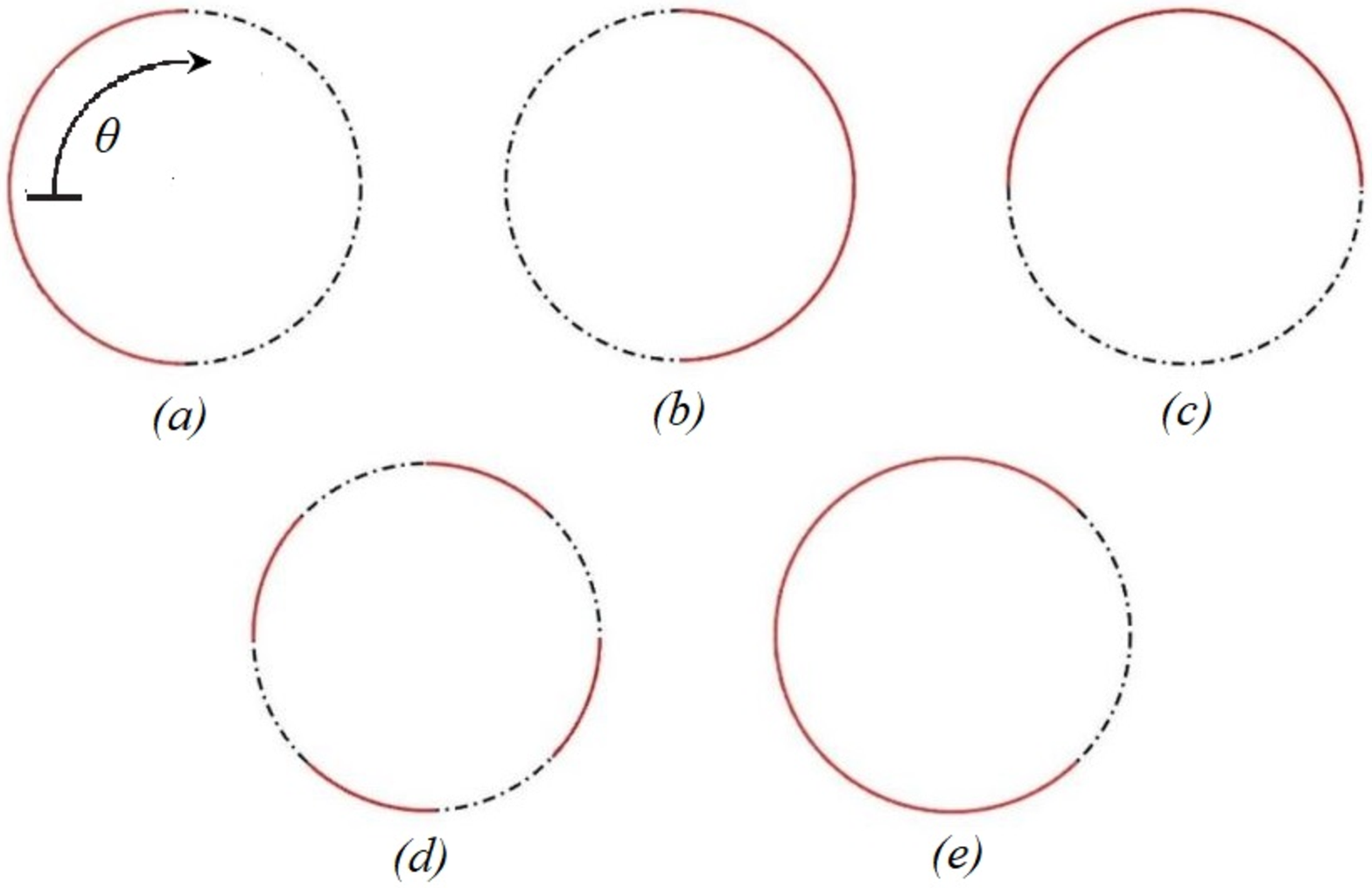}}
\caption{Schematic of partially superhydrophobic cylinder, case (a) front half, (b) rear half, (c) upper half, (d) $45^{\circ}$ slip/no-slip sections, and (e) $135^{\circ}$ section. Note that the solid and dashed lines represent the superhydrophobic and no-slip segments, respectively.}
\label{fig:Cases}
\end{figure}

Next, the characteristics of flow and heat transfer over partially superhydrophobic stationary cylinders are investigated. To proceed, five different cases have been considered which are depicted in Fig.~\ref{fig:Cases}$(a)$ to $(d)$, representing the application of slip over the front half, rear half, upper half, $45^{\circ}$ slip/no-slip sections and $135^{\circ}$ section of the cylinder, respectively.
The average values of $C_{d}$, $C_{lrms}$, $St$, $Nu$ and $Nu/C_{d}$ are reported in Tab.~\ref{tab:PartiallyData} for the aforementioned partially superhydrophobic cases. As can be seen, applying slip to the front and upper halves of the cylinder results in lower values of $C_{d}$ and $C_{lrms}$ compared to the rear half case. Also, the Strouhal number and $Nu/C_{d}$ attain larger values for the front and upper half cases. Furthermore, for the case of alternating slip/no-slip sections, i.e. case $(d)$, the values of flow and heat transfer parameters lie in between the front, upper, and rear half values. Lastly, it can be deduced that applying slip throughout a $135^{\circ}$ section of the cylinder surface leads to higher amounts of drag and lift reduction along with an increase in heat transfer rate. However, the value of $Nu$ is higher for the case of a fully superhydrophobic cylinder.
\begin{table}
\begin{center}
\caption{Values of the average drag coefficient, rms of lift coefficient, Strouhal number and Nusselt number for the five cases of partially superhydrophobic stationary cylinder. Note that each row corresponds to the cases mentioned in Fig.~\ref{fig:Cases}.}
\def~{\hphantom{0}}
\begin{tabular}{lcccccc}
\hline\hline
&   &  $C_{d}$   & $C_{lrms}$ & $St$ & $Nu$ & $Nu/C_{d}$ \\[3pt]\hline
\hline
& Case $(a)$ & $1.023$ & $0.281$ & $0.213$ & $25.754$ & $25.174$\\
\hline
& Case $(b)$ & $1.165$ & $0.335$ & $0.200$ & $19.820$ & $17.012$\\
\hline
& Case $(c)$ & $0.921$ & $0.321$ & $0.212$ & $23.201$ & $25.191$\\
\hline
& Case $(d)$ & $1.105$ & $0.304$ & $0.200$ & $22.420$ & $20.289$\\
\hline
& Case $(e)$ & $0.708$ & $0.102$ & $0.233$ & $26.990$ & $38.121$\\
\hline\hline
\end{tabular}
\label{tab:PartiallyData}
\end{center}
\end{table}

\begin{figure*}
\centerline{\includegraphics[width=\textwidth]{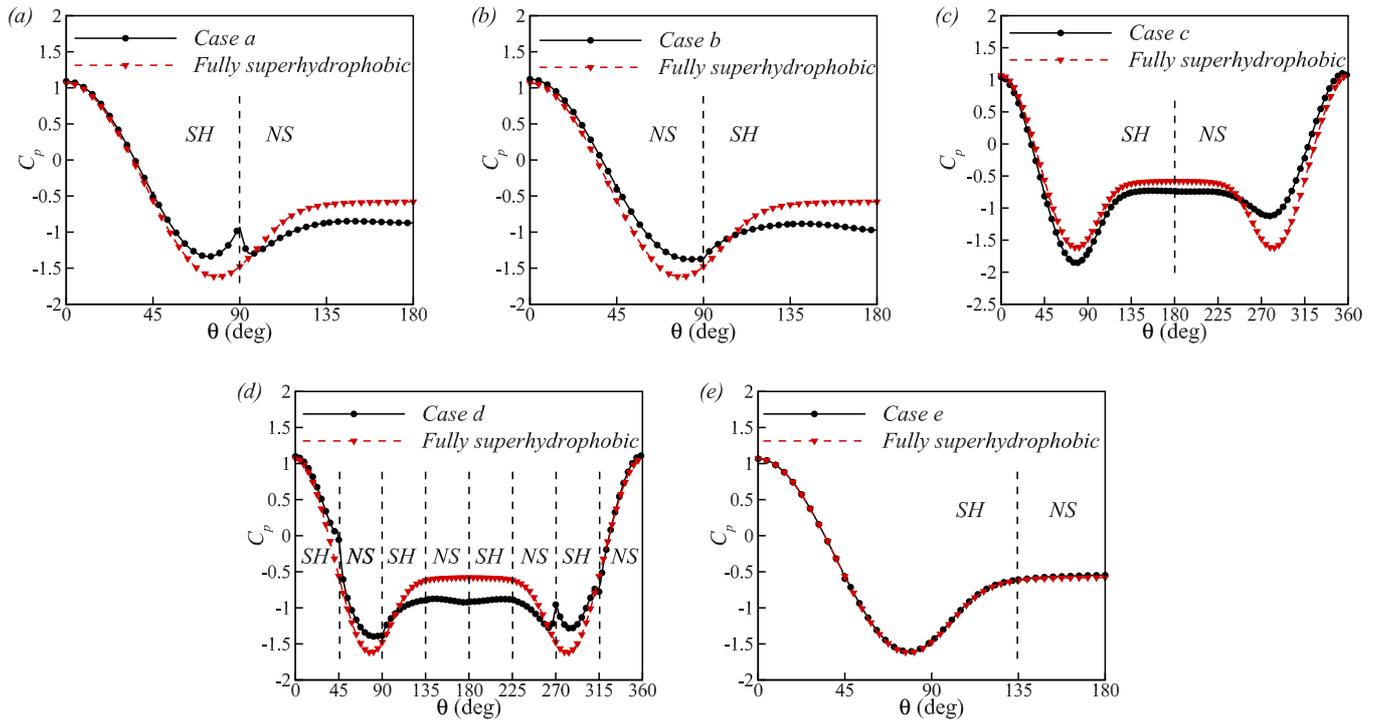}}
\caption{Local variation of the pressure coefficient for the five cases of partially superhydrophobic stationary cylinder, case (a) front half, (b) rear half, (c) upper half, (d) $45^{\circ}$ slip/no-slip sections, and (e) $135^{\circ}$ section. Note that NS and SH refer to the no-slip and superhydrophobic segments, respectively.}
\label{fig:CpPartially}
\end{figure*}

\begin{figure*}
\centerline{\includegraphics[width=\textwidth]{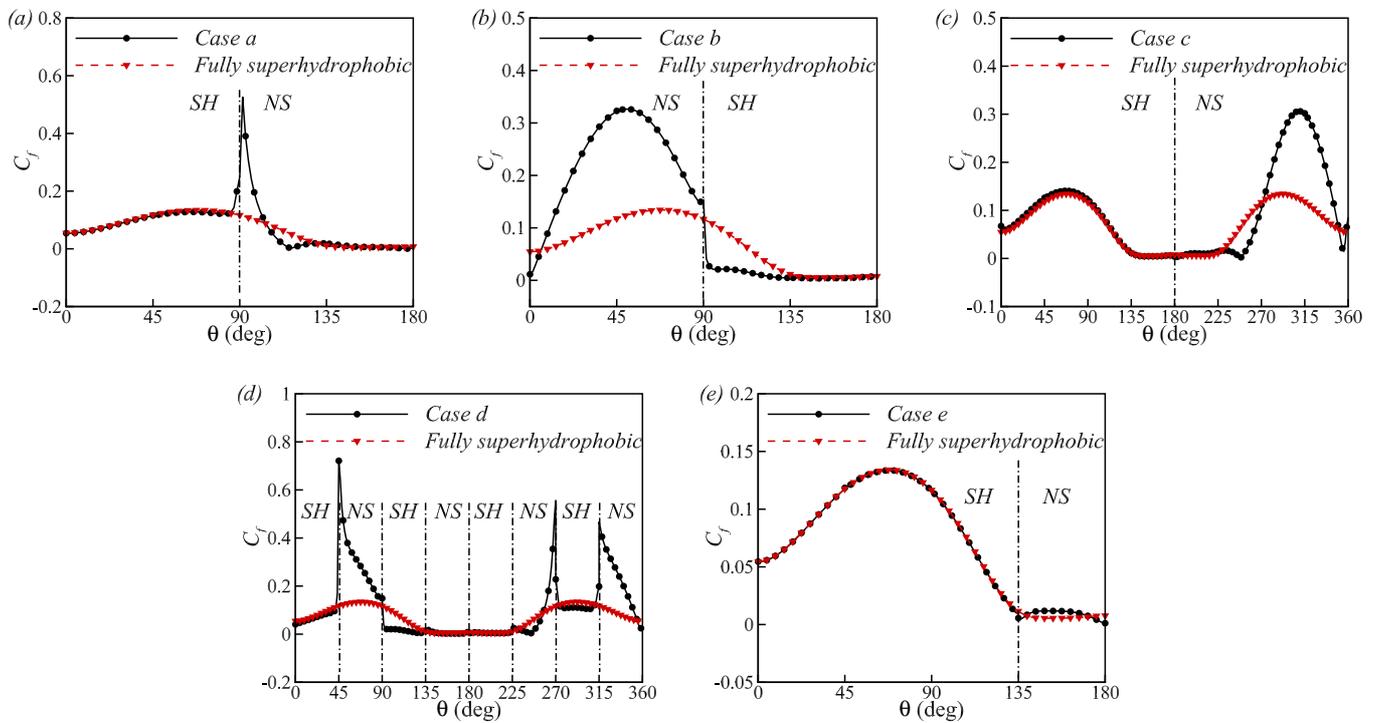}}
\caption{Local variation of the skin friction coefficient for the five cases of partially superhydrophobic stationary cylinder, case (a) front half, (b) rear half, (c) upper half, (d) $45^{\circ}$ slip/no-slip sections, and (e) $135^{\circ}$ section. Note that NS and SH refer to the no-slip and superhydrophobic segments, respectively.}
\label{fig:CfPartially}
\end{figure*}

In order to further analyze the previously mentioned trends, local distribution of the pressure and skin friction coefficients are displayed in Figs.~\ref{fig:CpPartially} and ~\ref{fig:CfPartially} for the five cases of partially superhydrophobic stationary cylinder, respectively. As is depicted, the variation of pressure coefficient is nearly the same for the cases of front and rear half slip. However, the skin friction coefficient is higher throughout the first quarter of the cylinder surface in case $(b)$, which results in a larger overall drag coefficient for this case. Regarding case $(c)$, the pressure coefficient is slightly lower than the case of a fully superhydrophobic cylinder up to $\theta = 225^{\circ}$. Afterwards, $C_{p}$ attains higher values in case $(c)$. Also, skin friction coefficient exhibits the same behavior as the fully superhydrophobic cylinder in a similar range of $\theta$, gaining larger values from $225^{\circ}$ onward. Alternating slip/no-slip sections, i.e. case $(d)$, leads to the appearance of several discontinuities and sudden jumps in the pressure and skin friction coefficient diagrams. These sudden changes are more pronounced in the first half of the cylinder, which is $\theta < 90$ and $\theta > 270$. However, shifting between the no-slip and slip conditions does not affect the behavior of $C_{p}$ and $C_{f}$ inside the $90 <\theta < 270$ range. Finally, case $(d)$ shows that the application of slip throughout a $135^{\circ}$ section of the cylinder results in the same trend of the pressure and skin friction coefficients, although the overall value of $C_{f}$ is marginally higher after $\theta = 135^{\circ}$ compared to the fully superhydrophobic cylinder.

\begin{figure*}
\centerline{\includegraphics[width=\textwidth]{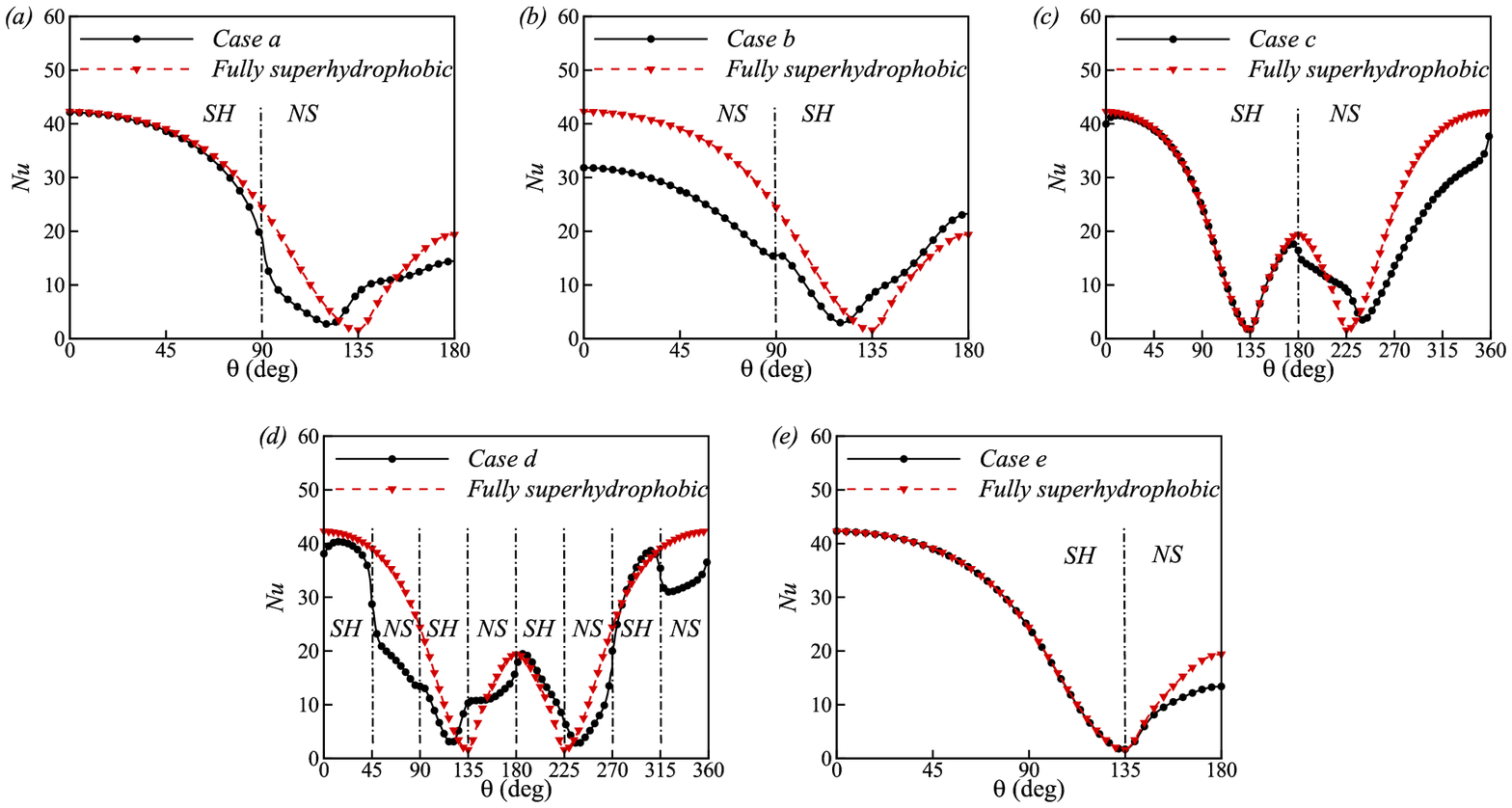}}
\caption{Local variation of the Nusselt number for the five cases of partially superhydrophobic stationary cylinder, case (a) front half, (b) rear half, (c) upper half, (d) $45^{\circ}$ slip/no-slip sections, and (e) $135^{\circ}$ section. Note that NS and SH refer to the no-slip and superhydrophobic segments, respectively.}
\label{fig:NuPartially}
\end{figure*}

Figure~\ref{fig:NuPartially} shows the distribution of Nusselt number over the cylinder surface for the aforementioned partially superhydrophobic cases. As can be seen, the overall value of $Nu$ is lower than that of the fully superhydrophobic cylinder in all of the no-slip sections, whereas the application of slip leads to improved heat transfer rates and higher Nusselt values.

\subsection{Transversely oscillating cylinder}
The effect of superhydrophobicity on the characteristics of flow and heat transfer over a transversely oscillating cylinder is analyzed in this section. First, the lock-in boundary for the case of a superhydrophobic oscillating cylinder is examined and compared to the results obtained for the no-slip case. It is found that the predominant frequency of the lift coefficient is equal to either the natural shedding frequency or the frequency of the cylinder oscillation. Additionally, for higher oscillation amplitudes, the lift coefficient generally displays multiple frequencies, one of which synchronizes with the cylinder oscillation frequency.

\begin{figure*}
\centerline{\includegraphics[width=\textwidth]{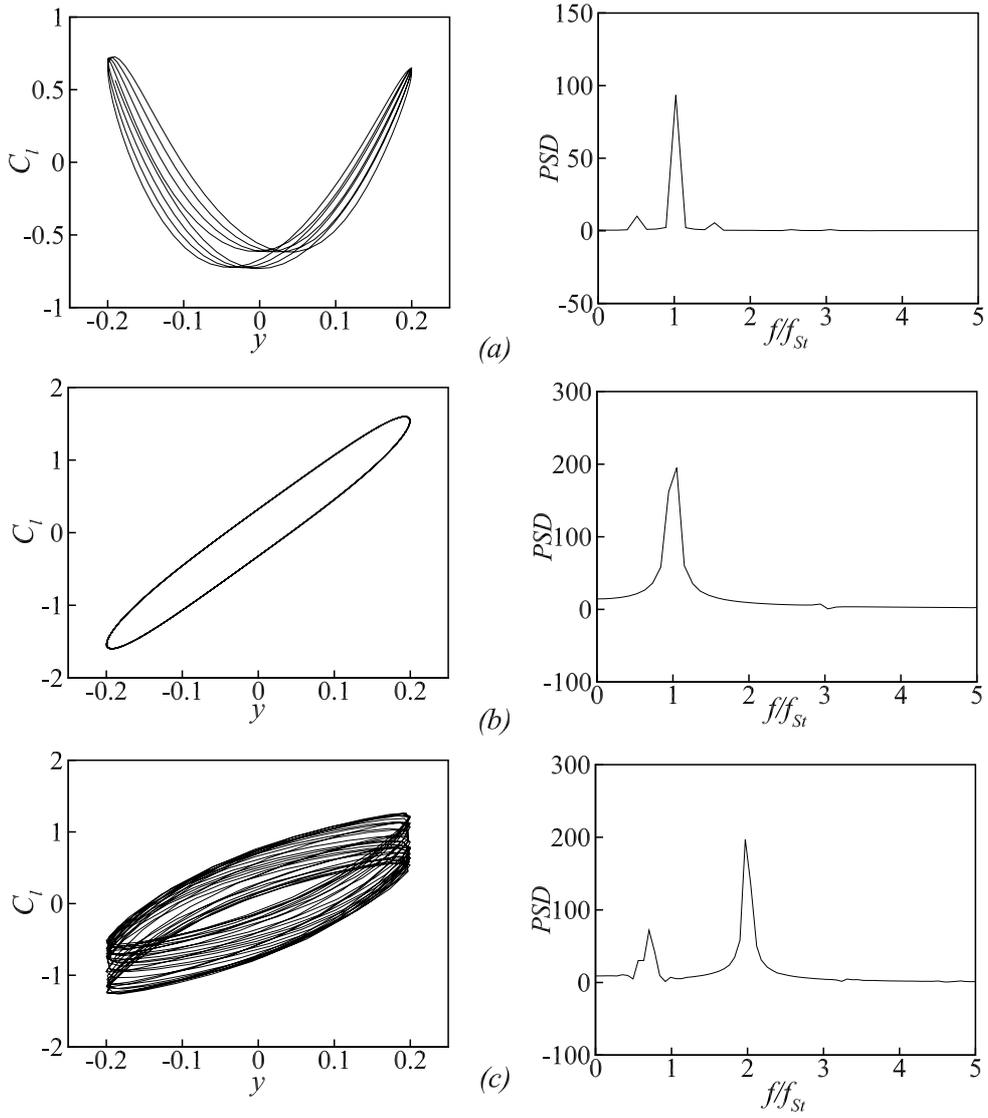}}
\caption{The Lissajous diagrams and corresponding FFT spectrum of (a) the no-slip cylinder at $F^*=0.5$, (b) and (c) the superhydrophobic cylinder at $F^*=1$ and $2$, respectively. $A^*=0.2$ in all cases.}
\label{fig:LisData}
\end{figure*}

In order to demonstrate the non-lock and the lock-in cases, three sets of Lissajous diagrams and their corresponding FFT spectrum have been shown in Fig.~\ref{fig:LisData}. It is well known that the variation of the lift coefficient against the displacement of the cylinder, i.e., the Lissajous plot, presents an irregular behavior for the non-lock cases, while a closed and regular pattern appears for the lock-in condition. This trend can be seen in Fig.~\ref{fig:LisData}$(a)$ when $F^*=0.5$ regarding the no-slip cylinder, and cases $(b)$ and $(c)$, regarding the superhydrophobic cylinder for the frequency ratios of 1 and 2, respectively. Also, the FFT spectrum of the non-lock case (Fig.~\ref{fig:LisData}$(a)$) indicates that the predominant frequency is equal to the natural shedding frequency, i.e., the Strouhal number. This means that the lock-in condition has not happened since the cylinder oscillation frequency is equal to half of the natural shedding frequency. On the other hand, due to the fact that the cylinder oscillation frequency is the same as the Strouhal number, the spectrum of data in Fig.~\ref{fig:LisData}$(b)$ shows a peak at the frequency ratio of 1, which means that lock-in has occurred. Figure~\ref{fig:LisData}$(c)$  depicts two peaks at the frequency ratios of around 1 and 2, meaning that both of the natural shedding oscillation and cylinder oscillation frequencies are present in the spectrum and therefore, synchronization is lost. It should be noted that $A^*=0.2$ for all three settings.

\begin{figure}
\centerline{\includegraphics[width=10cm]{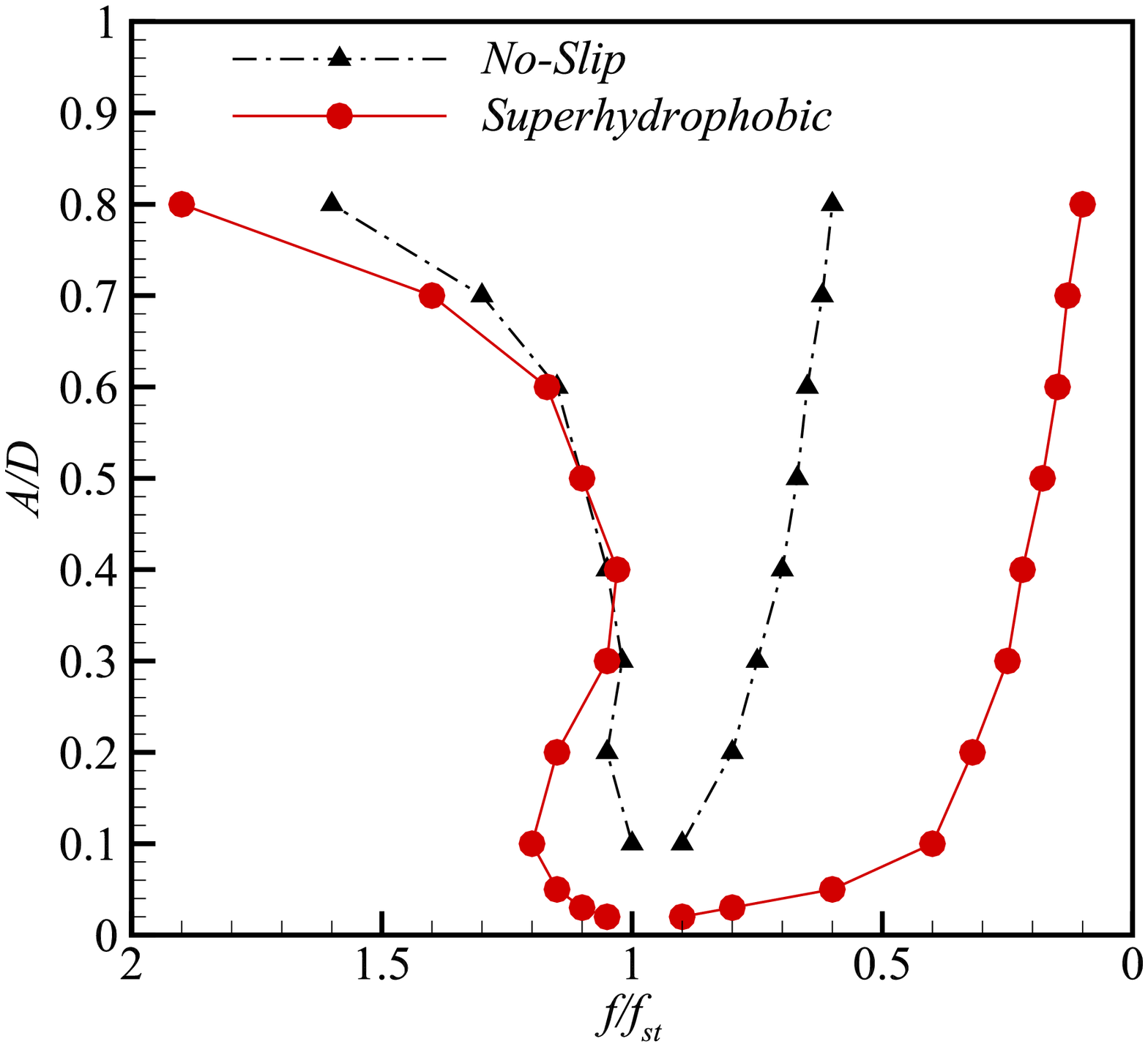}}
\caption{Comparison of the lock-in boundaries for the cases of the no-slip and superhydrophobic cylinder.}
\label{fig:LBResult}
\end{figure}

Fig.~\ref{fig:LBResult} presents boundaries of the lock-in occurrence for various amplitudes and frequencies regarding both the no-slip and slipped fields. It is seen that the lock-in boundary becomes notably wider on its low-frequency side. Superhydrophobicty also expands the high-frequency boundary in comparison to the no-slip case. Another feature to note is that the lowest oscillation amplitude for which the synchronization occurs goes down as a result of slippage. As mentioned before by Meneghini and Bearman~\cite{meneghini1995numerical}, this lowest amplitude is equal to 0.1 for the no-slip cylinder. However, this value is found to be around 0.02 for the superhydrophobic cylinder.

\begin{figure*}
\centerline{\includegraphics[width=\textwidth]{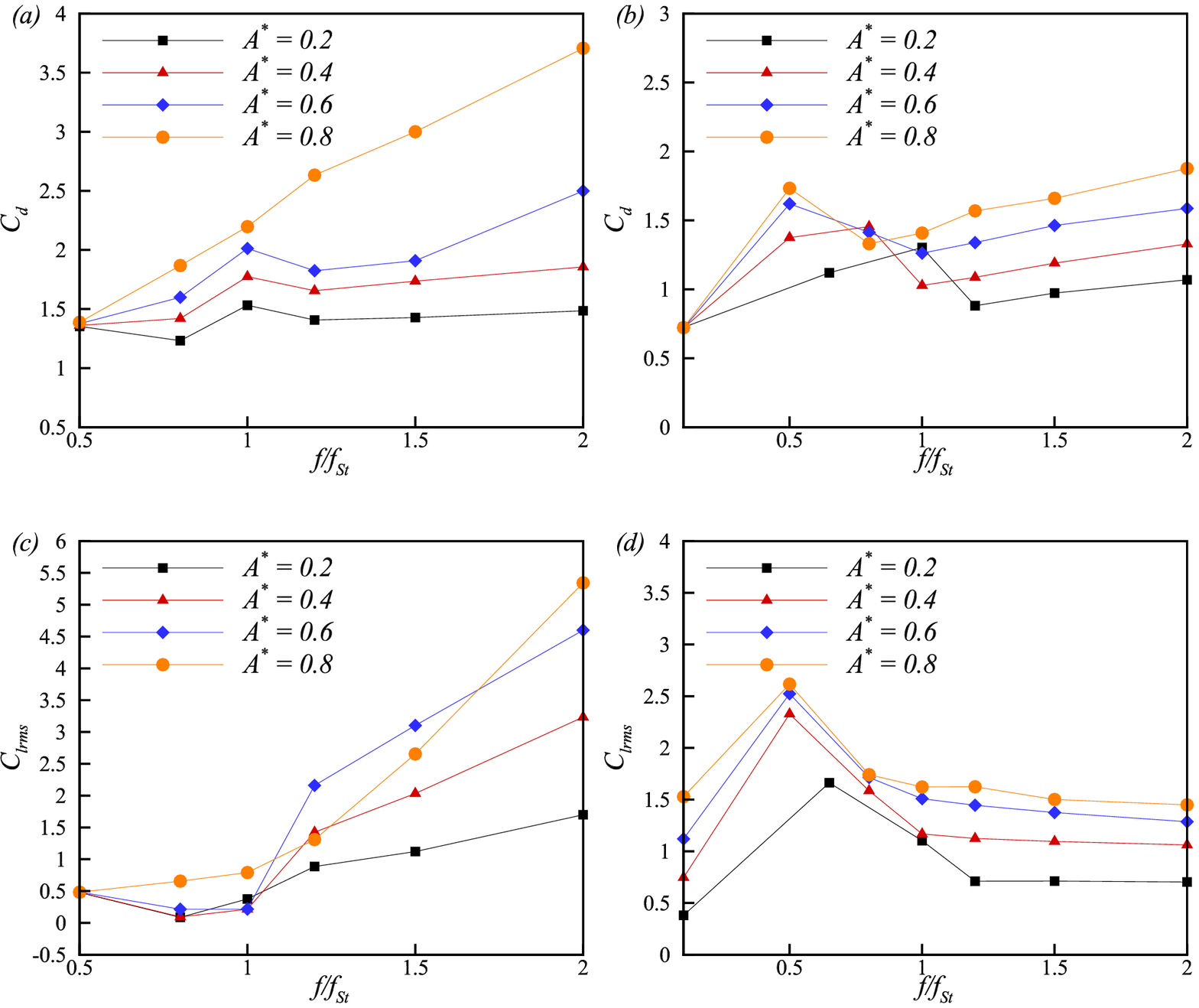}}
\caption{Comparison of $C_{d}$ and $C_{lrms}$ for the cases of no-slip cylinder (left column) and superhydrophobic cylinder (right column).}
\label{fig:CdClComp}
\end{figure*}

Variation of the mean drag coefficient for various oscillation frequencies and amplitudes has been shown in Fig.~\ref{fig:CdClComp}$(a)$ and $(b)$ for the cases of no-slip and superhydrophobic cylinders, respectively. Regarding the superhydrophobic case, for $A^*=0.2, 0.4, 0.6$, the value of average $C_d$ increases inside the lock-in region, and drops after reaching a maximum value. The same trend is initially followed by the no-slip oscillating cylinder, except that for the case of $A^*=0.8$, the mean drag coefficient continues to grow outside the synchronization range. It should be noted that the same behavior is not seen for $A^*=0.8$ in the case of the superhydrophobic cylinder. Therefore, it can be deduced that in the primary synchronization region, the oscillation amplitude for which the mean drag coefficient has an ever-increasing trend is shifted to higher values as a result of superhydrophobicity.
Figure~\ref{fig:CdClComp} also depicts the variation of $C_{lrms}$ for different oscillation amplitudes and frequencies in the primary lock-in range. As can be seen in Fig.~\ref{fig:CdClComp}$(c)$, the rms of lift coefficient decreases inside the synchronization range, and then attains higher values at higher oscillation frequencies for the no-slip cylinder. On the contrary, the superhydrophobic cylinder shows a completely reversed trend. As Fig.~\ref{fig:CdClComp}$(d)$ illustrates, super hydrophobicity increases the amount of $C_{lrms}$ inside the lock-in boundary and outside this range, the rms of the lift coefficient decreases.

\begin{figure*}
\centerline{\includegraphics[width=\textwidth]{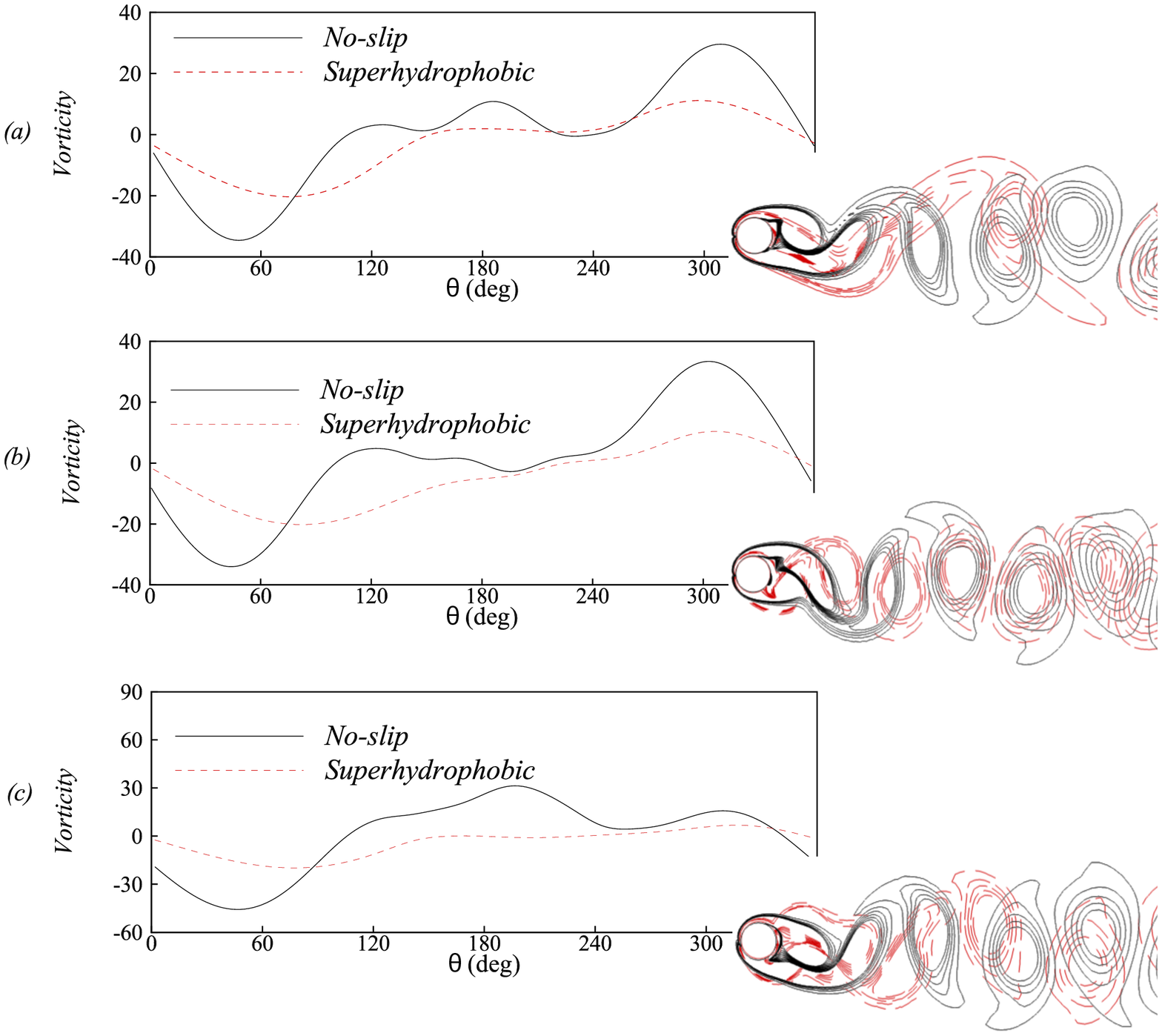}}
\caption{Local variation and the contours of vorticity at the oscillation amplitude of 0.2, (a) $F^*=0.5$, (b) $F^*=1$, and (c) $F^*=2$. In all frames, the cylinder is at its extreme upper position.}
\label{fig:LocalVorConComp}
\end{figure*}

In Figs.~\ref{fig:LocalVorConComp}$(a)$ to $(c)$, the vorticity contours have been shown for $A^*=0.2$ and frequency ratios of 0.5, 1, and 0.2, respectively. As can be seen, the vortices become more stretched for the case of a superhydrophobic cylinder at $F^*=0.5$ and are placed at a far more distance in the wake with respect to each other. At $F^*=1$, the slip causes the von Kármán vortex street to be more compressed and by increasing the frequency ratio up to 2, the vortices in the wake begin to part from each other again. Also, the shape of the vortices close to the cylinder surface differs at $F^*=2$, and depicts a higher level of attachment to the cylinder surface. Furthermore, the local distribution of vorticity on the surface of the cylinder has been shown in Fig.~\ref{fig:LocalVorConComp} for the same set of oscillation parameters. It is clear that superhydrophobicity suppresses the vorticity over the cylinder surface and the vorticity production occurs in the region $\theta=180^{\circ}-360^{\circ}$, while this range is around $120^{\circ}-240^{\circ}$ for the no-slip cylinder.

\begin{figure*}
\centerline{\includegraphics[width=\textwidth]{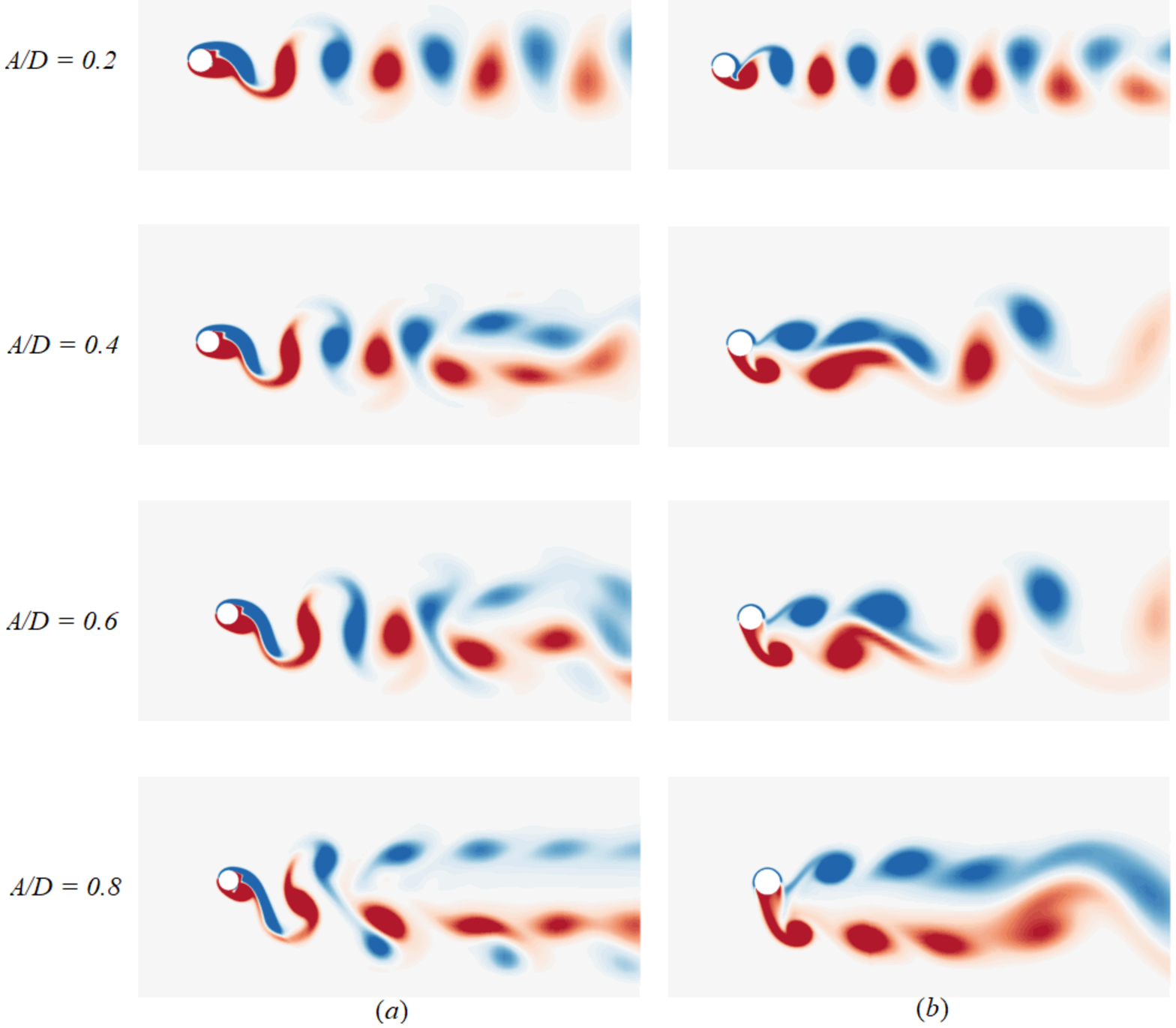}}
\caption{Comparison of the wake structure and vortex shedding modes for the cases of (a) the no-slip and (b) the superhydrophobic oscillating cylinder at $F^*=1$ and $A^*=0.2, 0.4, 0.6, 0.8$. In all frames, the cylinder is at its extreme upper position.}
\label{fig:wakeStOsc}
\end{figure*}

In accordance with the results of Leontini et al.~\cite{leontini2006wake}, the wake structure depends on the oscillation amplitude when the frequency of oscillation is around 1, resulting in appearance of the P+S vortex shedding mode at oscillation amplitudes of higher than 0.7. Figure~\ref{fig:wakeStOsc}$(a)$ and $(b)$ demonstrates the wake structure and the vortex shedding modes for the amplitudes of 0.2, 0.4, 0.6, 0.8 at $F^*=1$ for the cases of no-slip and superhydrophobic cylinder, respectively. The results of the no-slip case are in total agreement with those of Leontini et al.~\cite{leontini2006wake}. For the superhydrophobic cylinder, the transformation of the 2S vortex shedding mode to the P+S mode does not happen by crossing from $A^*=0.6$ to $0.8$. However, the vortex streets are formed near the top and the bottom sides of the cylinder moderately earlier in comparison to the no-slip case.

\begin{figure*}
\centerline{\includegraphics[width=\textwidth]{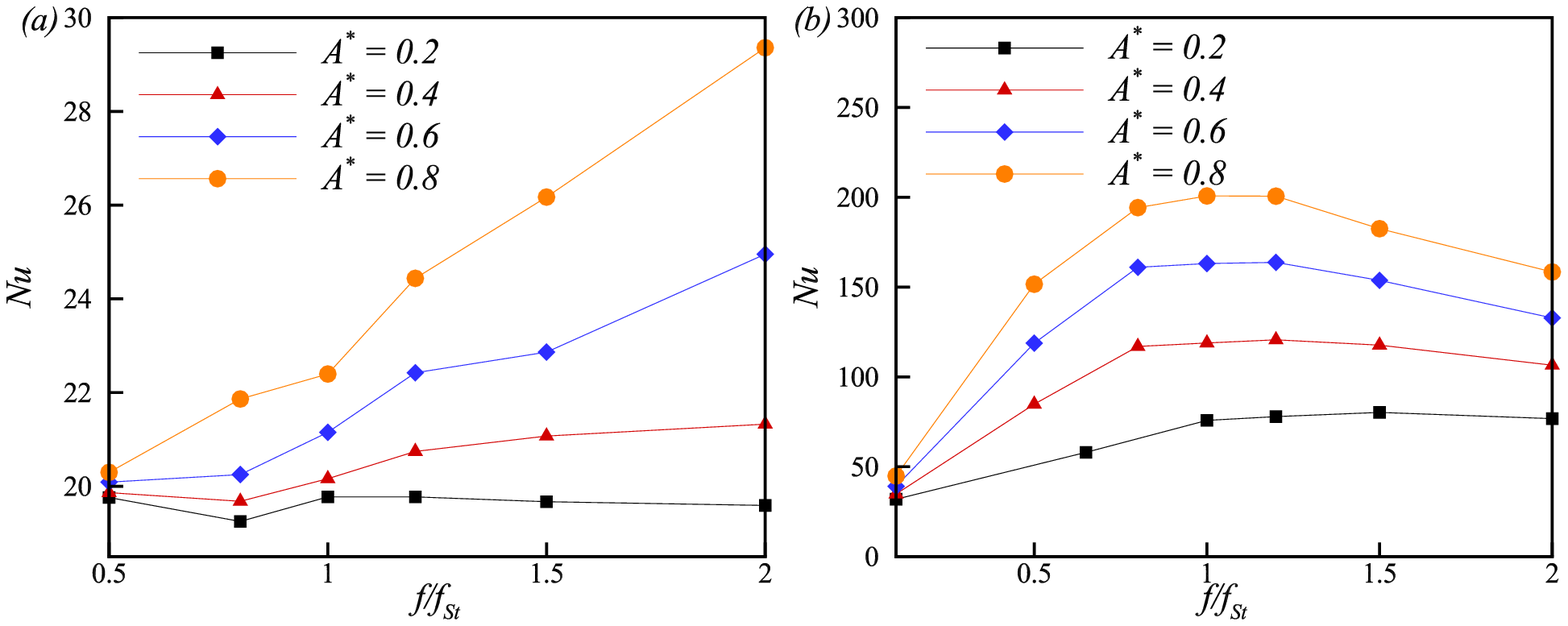}}
\caption{Comparison of the average Nusselt number for the cases of (a) the no-slip and (b) the superhydrophobic cylinder.}
\label{fig:NuComp}
\end{figure*}

Variation of the average Nusselt number for the cases of the no-slip and superhydrophobic cylinder has been demonstrated in Fig.~\ref{fig:NuComp}$(a)$ and (b), respectively. It is found that for the no-slip oscillating cylinder, the mean Nusselt number increases inside the lock-in range and becomes larger with increasing the oscillation amplitude. On the other hand, the superhydrophobic cylinder has the same increasing trend inside the synchronization boundary, but experiences a reduction as the oscillation frequency takes higher values. Also, the values of the mean Nusselt number are significantly higher than the results for the no-slip cylinder. For instance, the highest value of $Nu$ for the superhydrophobic cylinder is around 6.5 times greater than its counterpart for the no-slip case.

\begin{figure*}
\centerline{\includegraphics[width=\textwidth]{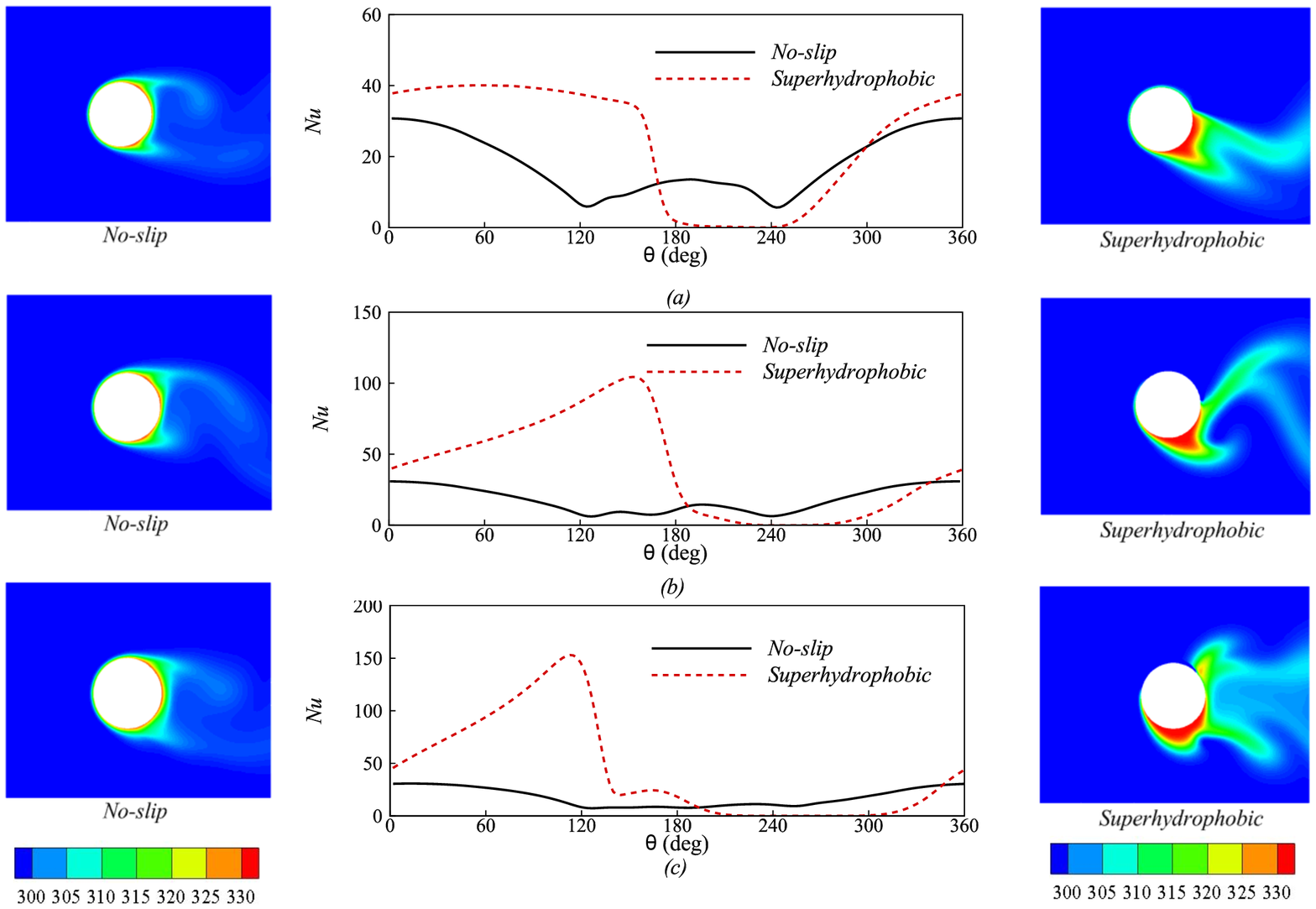}}
\caption{Local variation of the Nusselt number and temperature contours for the oscillation amplitude of 0.2, (a) $F^*=0.5$, (b) $F^*=1$, and (c) $F^*=2$. In all frames, the cylinder is at its extreme upper position.}
\label{fig:localNuTconComp}
\end{figure*}

Local distribution of the Nusselt number over the surface of the cylinder has been shown in Fig.~\ref{fig:localNuTconComp}$(a)$, $(b)$, $(c)$, at $A^*=0.2$ and $F^*=0.5, 1, 2$, respectively. The most significant fact we can detect here is that for the case of the no-slip cylinder, the Nusselt number decreases from a maximum magnitude at the front stagnation point, and attains its lowest value at around $\theta=120^{\circ}$. Then, after passing through a set of local maxima and minima in the separated region, it rises again to the same highest value near the rear stagnation point. However, regarding the superhydrophobic case, a completely different trend can be exhibited. Most importantly, the highest value of the Nusselt number no longer appears close to the front stagnation point. It is observed that the slip condition causes the Nusselt number to rise to its peak value before dropping to almost zero in the separated region.
Fig.~\ref{fig:localNuTconComp} also illustrates the temperature contours for the same set of parameters mentioned above. It can be seen that for the case of a superhydrophobic cylinder, due to the increased amount of the Nusselt number, a higher rate of heat is transferred into the flow wake. Furthermore, it is observed that at the region of $\theta=180^{\circ}-300^{\circ}$, the temperature gradient is almost zero and the uniform distribution of the temperature leads to the low values of $Nu$ mentioned before.

\begin{figure*}
\centerline{\includegraphics[width=0.9\textwidth]{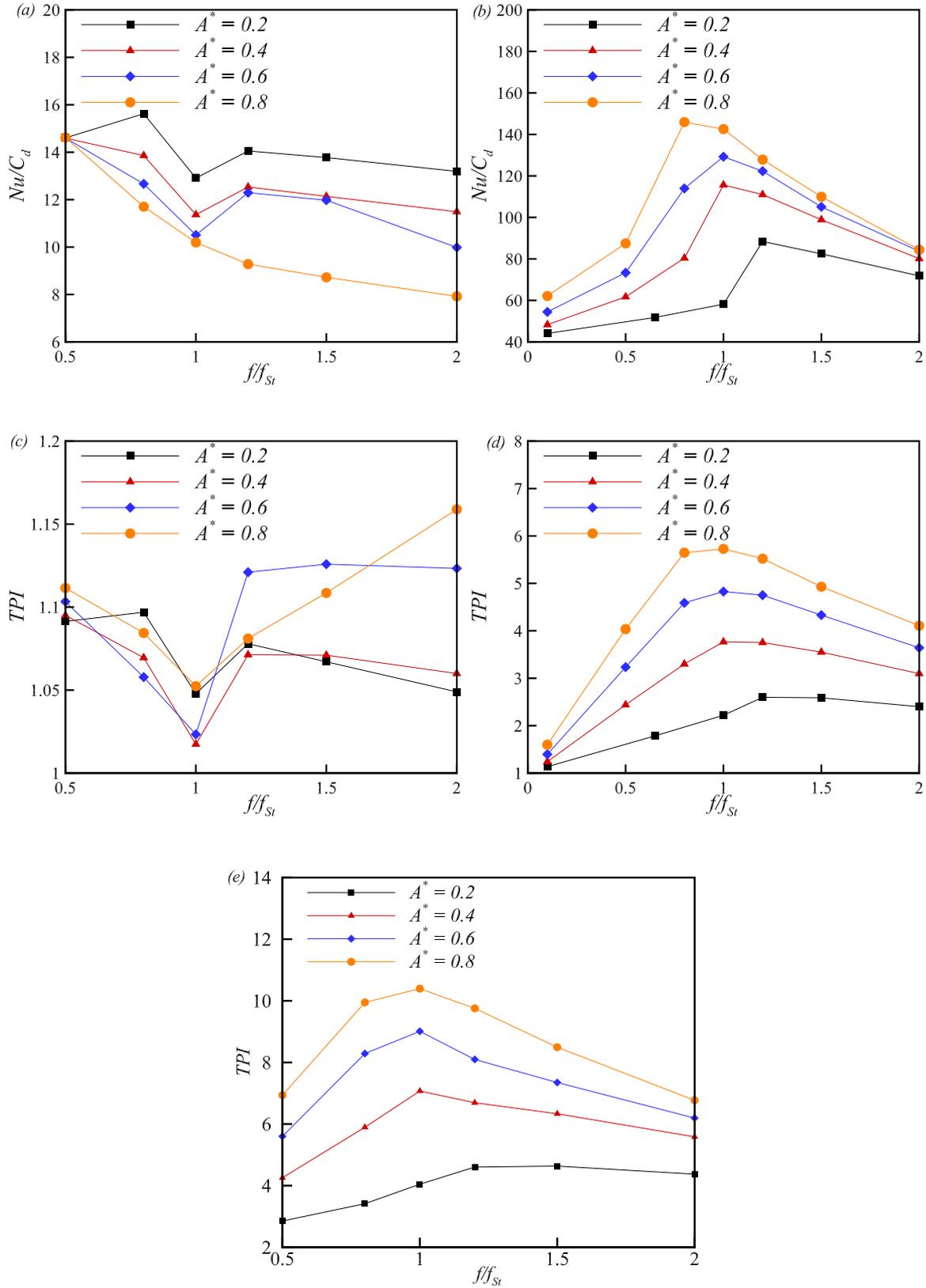}}
\caption{Comparison of the ratio of the Nusselt number and the drag coefficient for the cases of (a) the no-slip and (b) the superhydrophobic oscillating cylinders, and the thermal performance indices for the cases of (c) the no-slip, (d) and (e) the superhydrophobic oscillating cylinders. Regarding the TPI curves, note that the reference states are according to those of the stationary cylinder for (c) and (d), and the oscillating cylinder for (e).}
\label{fig:NuCdTPI}
\end{figure*}

Figure~\ref{fig:NuCdTPI}$(a)$ and $(b)$ show the variation of $Nu/C_{d}$ for the cases of the no-slip and superhydrophobic oscillating cylinders, respectively. The value of $Nu/C_{d}$ follows an overall decrease for the case of the no-slip cylinder, reaching a minimum at the lock-in state, i.e., $f/f_{St}=1$. This trend indicates that heat transfer enhancement can not be merely attributed to the variation of the Nusselt number, and the $Nu/C_{d}$ ratio provides a more precise analysis of heat transfer augmentation with respect to the changes of drag coefficient. On the other hand, as Fig.~\ref{fig:NuCdTPI}$(b)$ depicts, superhydrophobicity alters the trend of $Nu/C_{d}$ in a major way. Before the lock-in phenomenon occurs, the value of the Nusselt number divided by the drag coefficient is rising to a maximum, which is achieved at the lock-in state, and as the oscillation frequency takes higher values, $Nu/C_{d}$ decreases. Consequently, since applying the slip to the surface of the cylinder enhances the heat transfer rate while reducing the drag coefficient, it can be deduced that superhydrophobicity provides a better heat transfer performance for a wide range of oscillation amplitudes and frequencies.

The combined analysis of flow and thermal fields can be further extended by computing the thermal performance index (TPI) for our cases. The variation of TPI is shown in Fig.~\ref{fig:NuCdTPI}$(c)$ and $(d)$ for the no-slip and superhydrophobic cylinder, respectively. As equation~(\ref{eqTPI}) suggests, a reference state is needed to be defined in order to calculate TPI for each case. In Fig.~\ref{fig:NuCdTPI}$(c)$ and $(d)$, the values of $Nu$ and $C_{d}$ for the stationary cylinder have been considered as initial states. Thus, Fig.~\ref{fig:NuCdTPI}$(c)$ and $(d)$ represent the effect of oscillation on thermal performance of the cylinder for both cases of the no-slip and slipped cylinders. For the no-slip surfaces, the decreasing trend suggests that the increase of the Nusselt number for the oscillating cylinder can not cope with the higher values of the drag coefficient at the lock-in state. Therefore, TPI falls down to a minimum at the lock-in region. This trend supports the fact that the enhancement of heat transfer can not only be decided by an increase in the Nusselt number and the effect of the drag coefficient needs to be considered as well in applications with constant pumping power. However, as Fig.~\ref{fig:NuCdTPI}$(d)$ illustrates, the thermal performance index experiences an increase up to a maximum at the lock-in region. This feature is completely in contrast with that of the no-slip cylinder, showing a reducing trend. As a result, superhydrophobicity can enhance the thermal performance of the oscillating case by about 5 times due to the fact that it increases the Nusselt number and simultaneously reduces the drag coefficient.
The effect of superhydrophobicity on TPI has also been investigated for the oscillating cylinder in Fig.~\ref{fig:NuCdTPI}$(e)$. It should be noted that the reference states are defined as the value of the Nusselt number and the drag coefficient of the oscillating no-slip cylinder in this case. Similar to Fig.~\ref{fig:NuCdTPI}$(d)$, the thermal performance index attains a maximum at the lock-in condition, and exhibits ten times higher values in comparison to the data in Fig.~\ref{fig:NuCdTPI}$(c)$. This means that for the case of an oscillating cylinder, application of slip on the surface highly enhances the heat transfer from the cylinder compared to the no-slip state.

\section{Conclusions}
Two-dimensional laminar flow and heat transfer over superhydrophobic stationary and transversely oscillating cylinder is numerically studied. After the validation of results, the effect of slip on the mean flow and heat transfer characteristics of the stationary cylinder has been studied by means of analyzing the local distributions of various flow and heat transfer parameters. It is shown that increasing the amount of slip causes the vortex shedding in the wake to vanish at a specific threshold value. The mean drag coefficient, the amplitude of the lift coefficient and the magnitude of the rms lift coefficient reduce as a result of increased slip. The average Nusselt number is also shown to attain higher values for the case of a superhydrophobic cylinder for both the iso-temperature and iso-flux boundary conditions.

After choosing a fixed value for the slip coefficient, it is shown that superhydrophobicity reduces the total drag coefficient by 46.2 \% and increases the Nusselt number by 55 \%. Analysis of the local distributions of the pressure and skin friction coefficients resulted in the reduction of form and friction drag. Furthermore, the FFT spectrum of the lift coefficient reveals that the natural shedding frequency for the superhydrophobic cylinder is about 21 \% higher than that of the no-slip case. The effects of applying slip over different sections of the cylinder surface are also studied and the results show that the front and upper halves of the cylinder are relatively more effective in the reduction of force coefficients and heat transfer augmentation, with $135^{\circ}$ case being the optimum, resulting in a 47 and 85 \% decrease of the drag and lift amplitude coefficients. However, the fully superhydrophobic cylinder provides higher values of $Nu$, being around 5 \% higher than that of the $135^{\circ}$ case.

Regarding the transversely oscillating superhydrophobic cylinder, it is demonstrated that the boundary for the primary synchronization region is expanded on both the lower and higher frequency limits of the map. Furthermore, the average drag coefficient follows the same trend as the no-slip case for different amplitudes and frequencies of oscillation, except the point that the oscillation amplitude for which the mean drag continuously elevates with respect to the frequency ratio is shifted towards higher values. On the contrary, the trend of the rms lift coefficient for the superhydrophobic cylinder is completely different from the no-slip case, such that after rising to its maximum at the lock-in state, it goes down to lower values with increasing frequency ratio.

The wake structure is then analyzed and compared to the no-slip case, showing that the formation of the vortices on the cylinder surface and their diffusion into the wake has been significantly altered as a result of superhydrophobicity. Also, for the oscillation amplitudes and frequencies considered in this paper, the vortex shedding mode of P+S does not occur for the superhydrophobic cylinder. The investigation of the Nusselt number and its local distribution over the surface of the cylinder suggested that superhydrophobicity increases the heat transfer rate from the oscillating cylinder compared to the no-slip case. Further, the analysis of the ratio of the Nusselt number over the drag coefficient provides more insight on the relative importance of heat transfer enhancement with respect to drag reduction. In order to study this matter more thoroughly, an important variable called the thermal performance index (TPI) has been utilized and the effects of oscillation and slip have been analyzed on the variation of this parameter. For the case of the no-slip cylinder, it is shown that oscillation causes TPI to drop off to a minimum at the lock-in condition, whereas the superhydrophobic oscillating cylinder attains a peak value for TPI at this state. This means that oscillation has an enhancing effect on thermal performance of the superhydrophobic cylinder. The same feature was observed for the thermal performance of an oscillating superhydrophobic cylinder with respect to the no-slip oscillating case.


\end{document}